\documentclass[preprint,floatfix,11pt] {revtex4} 
 
\newcommand{\pvec}{\mathrm {\mathbf {p}}} 
\usepackage{graphicx}
\usepackage{subfigure}
\usepackage{xcolor}
\usepackage{amsmath}
\usepackage{enumerate}
\usepackage{tabularx}

\usepackage{color, soul}
\definecolor{darkblue}{rgb}{0,0,0.5}
\setulcolor{darkblue}

\begin{document}

\title{Analysis of Compton profile through information theory in H-like atoms inside impenetrable sphere}

\author{Neetik Mukherjee}
\altaffiliation{Email: neetik.mukherjee@iiserkol.ac.in.}

\author{Amlan K.~Roy}
\altaffiliation{Corresponding author. Email: akroy@iiserkol.ac.in, akroy6k@gmail.com.}
\affiliation{Department of Chemical Sciences
Indian Institute of Science Education and Research (IISER) Kolkata, 
Mohanpur-741246, Nadia, WB, India}

\begin{abstract}
Confinement of atoms inside various cavities has been studied for nearly eight decades. However, the Compton profile for such systems 
has not yet been investigated. Here we construct the Compton profile (CP) for a H atom radially confined inside a \emph{hard} 
spherical enclosure, as well as in \emph{free condition}. Some exact analytical relations for the CP's of circular or nodeless states
of free atom is presented. By means of a scaling idea, this has been further extended to the study of an H-like atom trapped 
inside an impenetrable cavity. The accuracy of these constructed CP has been confirmed by computing various momentum moments. 
Apart from that, several information theoretical measures, like Shannon entropy ($S$) and Onicescu energy ($E$) have been 
exploited to characterize these profiles. Exact closed form expressions are derived for $S$ and $E$ using the ground state
CP in free H-like atoms. A detailed study reveals that, increase in confinement inhibits the rate of dissipation of kinetic energy. 
At a fixed $\ell$, this rate diminishes with rise in $n$. However, at a certain $n$, this rate accelerates with progress in 
$\ell$. A similar analysis on the respective free counterpart displays an exactly opposite trend as that in confined system. 
However, in both free and confined environments, CP generally gets broadened with rise in $Z$. Representative calculations are 
done numerically for low-lying states of the confined systems, taking two forms of position-space wave functions: (a) 
exact (b) highly accurate eigenfunctions through a generalized pseudospectral method. In essence, CPs are reported for confined 
H atom (and isoelectronic series) and investigated adopting an information-theoretic framework.         
   
\vspace{10mm}

\end{abstract}
\maketitle

\section{introduction}
In quantum mechanics, electron density (ED) is the most important physical quantity and lies in the heart of chemistry. 
Because, it is inherently connected with the structure, bonding and reactions-the fundamental pillars. According to 
Hohenberg-Kohn theorem of density functional theory, when the exact density is known, any property of the system can be calculated 
exactly. It is larger near the nuclei and covalent chemical bonds. All kinds of chemical interactions affect ED. X-ray 
diffraction has provided the experimental information about ED. Similarly, electron momentum density (EMD) which is the 
momentum counterpart of ED, has also been used quite extensively to understand chemical systems. It can be extracted directly from 
$(e,2e)$ spectroscopy, position annihilation spectroscopy or X-ray Compton scattering (CS) \cite{compton23,lahtola11}.     

Compton effect is an inelastic scattering process \cite{compton23}, with use as spectroscopic probe in both single-particle 
excitation as well as in collective mode \cite{platzman65}. Linear CS for weekly bound electrons is well explained by the 
so-called impulse approximation \cite{eisen70}, where the electron is assumed to be quasi-free. {\color{blue}It means that, the 
binding energy of the electron is insignificant compared to the energy acquainted to it by the photon, so that the ultimate state 
of an electron may be adequately represented by a plane wave \cite{epstein73}.} Within such an approximation, one can write, 
\begin{equation}
\frac{\mathrm{d}^{2}\sigma}{\mathrm{d}\Omega \mathrm{d}\epsilon_{2}}=C(\epsilon_{1},\epsilon_{2},\phi) J(q). 
\end{equation}  
This equation indicates that, double differential cross section ($\sigma$) measures the quantity of photon scattered by 
matter having a solid angle $\Omega$ with energy $\epsilon_{2}$. Here, $\epsilon_{1}$ signifies the energy of incident photon
and $\phi$ the scattering angle. $C(\epsilon_{1},\epsilon_{2},\phi)$ depends on the experimental setup, while $J(q)$ corresponds
to the Compton profile (CP). It determines the projection of EMD along direction of scattering vector. Importantly it 
provides the intensity of Compton band. Finally, $q$ is the projection of target electron momentum upon scattering vector. 
The concept of CS is well known; for its useful properties and features, one can consult some of the excellent elegant reviews 
\cite{eisen70,cooper85,cooper04}. 

CS provides information about EMD. They can then be employed to estimate various momentum moments, $\langle p^{n}\rangle$. In atomic 
systems CP was utilized to compute momentum moments, which are directly connected to entropy optimization principle \cite{gadre79,
sears81}. It was also invoked to analyze isoelectronic atoms \cite{gadre79}. In last two decades, numerous attempts were made to 
adopt CPs in interpreting various physical and chemical properties in atoms, molecules and solids, both theoretically as well as 
experimentally \cite{son17}. 
The cutting-edge X-ray CS technique 
permits us to visualize the bonding in liquids \cite{okada15} and imaging the hole states of 
dopants in complex materials \cite{sakurai11}. 
Theoretical prediction of CP 
has also been a very fruitful and worthwhile research topic ever since the work of \cite{duncanson45}. The agreement between 
theory and experiment was improved by incorporating electron correlation in the wave function progressively accurately.  
Substantial amount of work has been reported using wave-function based techniques as well as density functional theory (DFT). A 
large set of closed-shell molecules were studied at Hartree-Fock (HF), various post-HF and DFT levels \cite{hakala04}. Role of 
basis set in this context is very critical and has been examined \cite{hart05}. Several such calculations adopting CI with 
singles excitation \cite{thakkar86}, singles-doubles excitations \cite{thakkar87} and multiple excitations (up to six fold) 
\cite{thakkar90} were reported. Further, CI calculation by perturbation with multi-configurational zeroth-order wave function via
iterative processes has been presented \cite{merawa90}. In DFT, the accuracy of CPs strongly depends on quality of 
exchange-correlation functional\cite{ragot06}.      

Many important concepts in physics such as electron correlation \cite{kubo05,pisani11}, EMD \cite{cooper71,cooper97,aguiar15}, 
Fermi surface determination \cite{isaacs99,dugdale06,wang10,koizumi11}, X-ray and $\gamma$-ray radiations \cite{bergstrom97,
porter08,phuoc12}, were probed through the help of CP. 
CP is successfully employed to understand the anisotropy in 
nature of hydrogen bond of crystalline ice \cite{isaacs00}. Similarly, the hydrogen bond signature in NH$_4$F$_9$ has been 
studied \cite{barbiellini09}. Interestingly, it can explain the metal-insulator transition in La$_{(2-2x)}$Sr$_{(1+2x)}$Mn$_2$O$_7$ 
\cite{barbiellini09a}. However, in spite of such wide range of applications, CP in \emph{confined} quantum systems has been rarely
investigated neither experimentally nor by theoretical methods {\color{blue}(pertaining a few examples)}, which we attempt to 
explore in this work.   

Ever since its inception, confinement of quantum system has emerged as a subject of topical interest in the field of physics, 
chemistry, biology, nano-science and nano-technology, attracting a large number of elegant books and review articles 
\cite{jaskolski96,dolmatov04,grochala07,sabin2009,sen2014electronic,koo18,mukherjee18,mukherjee18a}. Atoms, molecules constricted 
under cavities of varying size and shape, exhibit distinct fascinating changes in their physical and chemical properties from their 
free counterpart. Very recently, a new virial-like theorem has been proposed for these systems \cite{mukherjee19}. Extreme high 
pressure (of the order 
of multi-megabar) always influences almost all properties of a chemical system, including (i) the fate of a chemical reaction,
(ii) reduction in size of anion (iii) elongation of length in covalent bond (iv) increase in coordination number of an 
atom in a coordination complex, etc. At such high pressure range, new bond can be formed and existing ones gets deformed (usually 
shortened, but in certain cases, stretched too) \cite{grochala07}. Besides, the van der Waal's space gets 
compressed \cite{hemley00}. Atom under high pressure was first studied as early as in 1937 \cite{michels37}. Such a situation 
can be modeled by shifting the spatial boundary from infinity to a certain finite region. Depending upon the capacity 
of pressure one can simulate them by invoking two broad category of confining potentials, \emph{impenetrable (hard)} and 
\emph{penetrable (soft)}. {\color{blue} The effect of pressure on CP as well as on autocorrelation functions of MgO polymorphs were studied
in the framework of DFT employing periodic linear combination of atomic orbital method \cite{joshi12}. Interestingly, the experimental 
investigation of CS under high pressure was done before, for elemental silicon by utilizing synchrotron radiation and Mao-Bell 
version of the Merrill-Basset diamond anvil cell with a Be gasket up to a pressure of 20 Gpa. Moreover, the use of Laue monochromator 
and a special assembly of compound refractive lenses made this novel experimental setup successful. The detailed description about 
such unique establishment is available in \cite{tse05}.}       

In the last twenty years, emergence of information theoretical concept has provided a major impetus in many diverse field of 
science and technology \cite{sen12}. They characterize the single-particle density of a system (in conjugate $r$ and $p$ spaces) 
in different complementary ways. Arguably, these are the most eligible measures of uncertainty, as they do not make any 
reference to some given point of a corresponding Hilbert space \cite{bbi06}. Moreover, these are compactly connected to 
energetics and experimentally measurable quantities of a given system. Shannon entropy ($S$) is the arithmetic mean of 
uncertainty. {\color{blue}Onicescu energy ($E$) is the expectation value of density and it is also termed as second-order 
moment of density \cite{oni66}. $E$ is also called \emph{dis-equilibrium}, as it measures the deviation of a distribution from 
\emph{equilibrium} \cite{shiner99}. Importantly,} being reciprocally  connected to $S$, $E$ usually upholds the inferences obtained 
from $S$.  {\color{blue} Particularly, in a given space, the increase of spreading in a density distribution is quantified by an 
increment in $S$ and decay in $E$.} Both $S,E$ are successfully employed to quantify various density-distributions produced from 
relevant theoretical or experimental processes. {\color{blue} $S$ has connection to Colin conjecture \cite{remirez97,site15}, 
atomic avoided crossing \cite{he15}, electron correlation effect \cite{site15}, configuration interaction \cite{alcoba16}, 
quantum entanglement in artificial atom \cite{nagy06b, amovilli04}, bond formation \cite{nalewajski08}, elementary chemical 
reactions \cite{rosa10}, orbital-free DFT \cite{nagy15}, 
aromaticity \cite{noor10} etc. Some of these like the avoided crossing occurs under confinement condition. 
Further, $S$ has its distinct ability to characterize artificial quantum systems designed by placing an atom or molecule 
inside a foreign environment \cite{sen2014electronic}. 
Likewise, $E$ has been widely employed to investigate the correlation energy and first ionisation potential 
in atomic and molecular systems \cite{florres16}. In stressed condition, the change in confinement strength leads to the variation 
in electron density distributions. This effect can be characterized by employing these information measures.}
 

In this article, we intend to show the effect of high pressure on CP of a confined quantum system considering a H-like atom 
under a rigid impenetrable well {\color{blue}(solid hydrogen)}, as test case. {\color{blue} The \emph{first principles} study of an 
atom trapped inside a fullerene cavity can provide accurate results. But they may not direct us to a simple interpretation of the calculated 
properties \cite{jasko96}. The cavity model has been designed using experimental results \cite{dolmatov12}. In the present 
scenario, the motive is to extract a qualitative idea about the impression of high pressure on CPs of CHA. To a certain extent, 
this can serve the purpose}. The designed CPs are characterized by employing a few information-theoretic measures, 
which helps uncover the effect of confinement on CP. These measures will act as descriptor in interpreting the CPs. This 
investigation will act as a threshold to discern the influence of various confined environments (such as quantum dot, encapsulated 
atom in fullerene cavity \cite{dolmatov12} and so on) on CPs. It is worthwhile noting that, CP for a free H atom (FHA) was studied before 
by some researchers \cite{duncanson45,epstein73,cooper85,cooper04}, however, we are not aware 
of any similar work in a confined H atom (CHA). Besides, the information analysis of CP in either FHA or CHA is not reported 
as yet. Therefore, at first, we examine the CP in an FHA; it is found to be possible to obtain a generalized expression for 
\emph{circular} {\color{blue}$(n-l=1)$} states, while for other states it needs to be calculated numerically. In the next step, we 
analyze the Compton Shannon entropy ($S^c$), Compton  
Onicescu energy ($E^c$) and generalized entropic moment ($\alpha$ order) using the Compton density, in various states in FHA. As it 
turns out, these quantities can be derived in closed form \emph{only for the lowest state} of FHA. Now to pursue the above relevant 
quantities 
in a CHA, we apply the following form of potential: $v_{c}=\infty$ at $r \ge r_{c},$ and \emph{zero} elsewhere. Here, $v_c$ represents 
the perturbing potential and $r_{c}$ the confining radius. The calculated CPs were tested by evaluating several momentum moments. 
It is well established that one can calculate the expectation value of any of the desired momentum moments 
in a given state, starting from their respective CP \cite{epstein73}. 
Later, this entire idea has been envisaged to confined H-like atoms (with varying $Z$, the nuclear charge) by introducing a novel 
scaling relation. For our calculations, we have chosen six states corresponding to $n \leq 3$, which suffices the current purpose. 
These are done by means of two wave functions, \emph{viz}, \emph{exact} wave function available in terms of 
Kummer hypergeometric function and the accurate wave function obtained through a generalized pseudospectral (GPS) method. It is  
verified that the results of these two methods are practically identical. The article is organized as follows. Section~II presents a 
brief summary of various aspects of methodology used in this work. Section~III provides an in-depth discussion of the results 
for both FHA and CHA. Finally we conclude with a few remarks and future prospects, in Sec.~IV.          

\section{Methodology}
\subsection{Theoretical formalism}
{\color{blue} The radial Schr\"odinger equation for a spherically confined system is expressed as, 
\begin{equation}\label{eq:1a}
	\left[-\frac{1}{2} \ \frac{d^2}{dr^2} + \frac{l (l+1)} {2r^2} + v(r) +v_c (r) \right] \psi_{n,l}(r)=
	\mathcal{E}_{n,l}\ \psi_{n,l}(r),
\end{equation}
where $v(r)=-Z/r \ (Z=1$ for H atom). Our desired high pressure confinement is established by 
invoking the potential: $v_c(r) = +\infty$ for $r > r_c$, and 0 for $r \leq r_c$, where $r_c$ implies the radius of 
the box. This equation needs to be solved under Dirichlet boundary condition, $\psi_{n,l} (0)=\psi_{n,l}(r_c)=0$.}
The \emph{exact} wave function for CHA {\color{blue}is obtained by solving Eq.~(\ref{eq:1a})}, which is expressible in terms of 
Kummer's M function (confluent hypergeometric) \cite{burrows06},   
\begin{equation}\label{eq:1}
\psi_{n, \ell}(r)= N_{n, \ell}\left(2r\sqrt{-2\mathcal{E}_{n,\ell}}\right)^{\ell} \ _{1}F_{1}
\left[\left(\ell+1-\frac{1}{\sqrt{-2\mathcal{E}_{n,\ell}}}\right),(2\ell+2),2r\sqrt{-2\mathcal{E}_{n,\ell}}\right] 
e^{-r\sqrt{-2\mathcal{E}_{n,\ell}}},
\end{equation}
with $N_{n, \ell}$ denoting normalization constant and $\mathcal{E}_{n,\ell}$ corresponding to energy of a state represented by 
$n,\ell$ quantum numbers. At $r=r_{c}$, this equation becomes \emph{zero}. It is a transcendental type equation and becomes 
useful when $\mathcal{E}_{n,\ell}$ are known. At $r=r_{c}$,  
\begin{equation}
_{1}F_{1}\left[\left(\ell+1-\frac{1}{\sqrt{-2\mathcal{E}_{n,\ell}}}\right),(2\ell+2),2r_{c}\sqrt{-2\mathcal{E}_{n,\ell}}\right] = 0.
\end{equation}

For a certain $\ell$, first root confirms the energy of the lowest-$n$ state ($n_{lowest}=\ell+1$), with successive roots signifying 
excited states. It is instructive to mention 
that, to construct the exact wave function of CHA for a definite state, one needs to provide energy eigenvalue of that state. 
In current purpose, $\mathcal{E}_{n,\ell}$ are 
computed by means of the GPS method, which has produced highly accurate eigenvalues for a 
number of central potentials, in both free and confined condition \cite{roy04,roy05,roy13}. 

In this present communication, our objective is to construct the spherically averaged EMD, which is the foundation to generate 
the CPs. In order to do that, at first, the $p$-space wave function can be obtained numerically by the following standard equation,  
\begin{equation}
\begin{aligned}
\psi_{n,\ell}(p) & = & \frac{1}{(2\pi)^{\frac{3}{2}}} \  \int_0^\infty \int_0^\pi \int_0^{2\pi} \psi_{n,\ell}(r) \ \Theta(\theta) 
 \Phi(\phi) \ e^{ipr \cos \theta}  r^2 \sin \theta \ \mathrm{d}r \mathrm{d} \theta \mathrm{d} \phi,  \\
      & = & \frac{1}{2\pi} \sqrt{\frac{2\ell+1}{2}} \int_0^\infty \int_0^\pi \psi_{n,\ell} (r) \  P_{\ell}^{0}(\cos \theta) \ 
e^{ipr \cos \theta} \ r^2 \sin \theta  \ \mathrm{d}r \mathrm{d} \theta.  
\end{aligned}
\end{equation}
Note that, $\psi(p)$ needs to be normalized. Integrating over $\theta$ and $\phi$ variables, this equation can be further modified to, 
\begin{equation}
\psi_{n,\ell}(p)=(-i)^{\ell} \int_0^\infty \  \frac{\psi_{n,\ell}(r)}{p} \ f(r,p)\mathrm{d}r.    
\end{equation}
Depending on $\ell$, this can be expressed in following simplified form ($m'$ starts with 0),  
\begin{equation}
\begin{aligned}
f(r,p) & = & \sum_{k=2m^{\prime}+1}^{m^{\prime}<\frac{\ell}{2}} a_{k} \ \frac{\cos pr}{p^{k}r^{k-1}} +  
            \sum_{j=2m^{\prime}}^{m^{\prime}=\frac{\ell}{2}} b_{j} \ \frac{\sin pr}{p^{j}r^{j-1}}, \ \ \ \ \mathrm{for} \ 
            \mathrm{even} \ \ell,   \\
f(r,p) & = & \sum_{k=2m^{\prime}}^{m^{\prime}=\frac{\ell-1}{2}} a_{k} \ \frac{\cos pr}{p^{k}r^{k-1}} +  
\sum_{j=2m^{\prime}+1}^{m^{\prime}=\frac{\ell-1}{2}} b_{j} \ \frac{\sin pr}{p^{j}r^{j-1}}, \ \ \ \ \mathrm{for} \ \mathrm{odd} \ \ell.
\end{aligned} 
\end{equation}
The coefficients $a_{k}$, $b_{j}$ of even-$\ell$ and odd-$\ell$ states are obtained analytically. In order to get further details, 
please see the Tables~I and II in reference \cite{mukherjee18a}. 

The spherically averaged EMD or the mean radial distribution function for a definite $n,\ell$-state in $p$ space, $I_{n,\ell}(p)$, 
can be extracted as \cite{epstein73,gadre79},
\begin{equation}
I_{n,\ell}(p)\mathrm{d}p=\int_{\omega} \left[\psi_{n,\ell}(\pvec) \psi_{n,\ell}^{*}(\pvec)p^{2}\right]\mathrm{d}\pvec 
= \psi_{n,\ell}(p) \psi_{n,\ell}^{*}(p)p^{2} \mathrm{d}p,
\end{equation} 
where $\omega$ is an element of solid angle for $p$. In a given state, $I_{n,\ell}(p)\mathrm{d}p$ signifies the probability 
that $p$ has a magnitude between $p$ and $p+dp$. Therefore, $\int_{0}^{\infty} I_{n,\ell}(p) \mathrm{d}p=1$. Now, assuming the 
impulse approximation, the desired spherically averaged CP of a CHA can be expressed as follows, 
\begin{equation}\label{eq:15}
J_{n,\ell}(q)=\int_{|q|}^{\infty} \frac{I_{n,\ell}(p)}{p} \mathrm{d}p.  
\end{equation}
$J_{n,\ell}(q)$ is a function merely of $q$, with its peak at $q=0$, while $q$ is the projection of target-electron momentum 
upon the scattering vector. The momentum moments are then given by,
\begin{equation}\label{eq:16}
\begin{aligned}
\left \langle \frac{1}{p} \right\rangle_{n,\ell} = 2J_{n,\ell}(q=0),  \ \ \ \
\langle p^{m} \rangle_{n,\ell} = 2(m+1) \int_{0}^{\infty} q^{m}J_{n,\ell}(q) \mathrm{d}q.
\end{aligned}
\end{equation} 
Here, our interest is to estimate $S, E$ for CPs; these can be easily expressed in terms of $J(q)$ as,
\begin{equation}\label{eq:14}
\begin{aligned}
S^{c}_{n,\ell}=-\int J_{n,\ell}(q) \ln J_{n,\ell}(q) \mathrm{d}q, \ \ \ \ \ \ E^{c}_{n,\ell}= \int J_{n,\ell}^{2}(q) \mathrm{d}q. 
\end{aligned}
\end{equation} 
A deeply bound electron has a very flat and broad momentum distribution \cite{eisen70}. As a consequence, its CP is also 
broad. This broadness of distribution can be quantified by $S_{c}, E_{c}$. Hence, these measures can act as descriptor about the 
bound effect on an electron within a quantum system. Our current study will convincingly establish this interpretation.  

\subsection{CHA isoelectronic series}
In case of the isoelectronic series of CHA, it is interesting to investigate the influence of $Z$ on CPs. In this sub-section, 
analytical relations between CP and $Z$ have been established by using some scaling properties. The required radial Schr\"odinger 
equation, in terms of a constrained coulomb potential, can be written as,
\begin{equation}\label{eq:2}
\begin{aligned}
-\frac{\hbar^{2}}{2m} \nabla^{2} \psi_{n,\ell}(r) -\frac{Z}{r} \psi_{n,\ell}(r) +V_{0}\theta(r-r_{c}) \psi_{n,\ell}(r) = 
\mathcal{E}_{n,\ell} \psi_{n,\ell}(r), \\
\theta(r-r_{c})=0  \ \ \ \mathrm{at} \ \ \ r \leq r_{c},  \ \ \ \ \ \ \theta(r-r_{c}) =1 \ \ \ \mathrm{at} \ \ \ r > r_{c}. 
\end{aligned}  
\end{equation}
Here $\theta(r-r_{c})$ is the Heaviside theta function and $V_{0}$ is taken to be an infinitely large positive number. After doing 
some straightforward mathematical manipulations, by transforming $r= \lambda r_{0}$ and assuming $\lambda = \frac{\hbar^{2}}{mZ}$ 
one leads to the following form of Hamiltonian,   
\begin{equation}\label{eq:3}
-\frac{1}{2} \nabla_{0}^{2} \psi_{n,\ell}(r_{0}) -\frac{1}{r_{0}} \psi_{n,\ell}(r_{0}) + V_{0}\theta\left(r_{0}-
\frac{mZr_{c}}{\hbar^{2}}\right) \psi_{n,\ell}(r_{0}) =
\frac{\hbar^{2}}{mZ} \mathcal{E}_{n,\ell} \psi_{n,\ell}(r_{0}). 
\end{equation}   
where $r_{0}$ is the new variable with dimension M$^{-1}$L$^{-3}$T$^{3}$I. We assume $\frac{\hbar^{2}}{mZ^{2}}V_{0} \approx V_{0}$ 
as $V_{0}$ approaches to $\infty$. Now, a comparison of Eqs.~(\ref{eq:2}) and (\ref{eq:3}) yields,
\begin{equation}
\begin{aligned}
\mathcal{E}_{n,\ell} \left(\frac{\hbar^{2}}{m}, Z, r_{c}\right) = \frac{mZ^{2}}{\hbar^{2}} \mathcal{E}_{n,\ell} \left(1, 1, 
\frac{mZr_{c}}{\hbar^{2}}\right), \\
\psi_{n,\ell} \left(\frac{\hbar^{2}}{m}, Z, r_{c}, r\right) = A \psi_{n,\ell} \left(1, 1, \frac{mZr_{c}}{\hbar^{2}}, r_{0}\right).
\end{aligned}
\end{equation}   
Applying the normalization {\color{blue}condition}, we obtain $A =\lambda^{-\frac{3}{2}}$. The $p$-space counterpart of $\psi_{n,l}$ can be 
achieved by performing the Fourier transformation. It is important to mention that, $p= \frac{p_{0}}{\lambda}$. 
\begin{equation}
\phi_{n,\ell} \left(\frac{\hbar^{2}}{m}, Z, r_{c}, p\right) = \lambda^{\frac{3}{2}} \phi_{n,\ell} \left(1, 1, 
\frac{mZr_{c}}{\hbar^{2}}, p_{0}\right).
\end{equation} 
Therefore, the spherically averaged momentum density ($I_{n,\ell}$) will take the form,
\begin{equation}\label{eq:4}
\begin{aligned}
I_{n,\ell} \left(\frac{\hbar^{2}}{m}, Z, r_{c}, p\right)p^{2} \ \mathrm{d}p & = \phi_{n,\ell} \left(\frac{\hbar^{2}}{m}, Z, r_{c}, 
p\right)
\phi_{n,\ell}^{*} \left(\frac{\hbar^{2}}{m}, Z, r_{c}, p\right) p^{2} \  \mathrm{d}p \\
& = \lambda^{3} \phi_{n,\ell} \left(1, 1, \frac{mZr_{c}}{\hbar^{2}}, p_{0}\right) \phi_{n,\ell}^{*} \left(1, 1, \frac{mZr_{c}}
{\hbar^{2}}, p_{0}\right) 
\left(\frac{p_{0}}{\lambda}\right)^{2} \ \frac{\mathrm{d}p_{0}}{\lambda}\\ 
& = I_{n,\ell} \left(1, 1, \frac{mZr_{c}}{\hbar^{2}}, p_{0}\right)p_{0}^{2} \ \mathrm{d}p_{0}. 
\end{aligned} 
\end{equation}   
Some further manipulation leads to the expression, 
\begin{equation}\label{eq:5}
\begin{aligned}
\langle p^{m} \rangle_{n,\ell} & = \int p^{m} I_{n,\ell} \left(\frac{\hbar^{2}}{m}, Z, r_{c}, p\right)p^{2} \ \mathrm{d}p \\
& = \int \left(\frac{p_{0}}{\lambda}\right)^{m} I_{n,\ell} \left(1, 1, \frac{mZr_{c}}{\hbar^{2}}, p_{0}\right)p_{0}^{2} \ 
\mathrm{d}p_{0} \\
& = \frac{1}{\lambda^{m}} \int p_{0}^{m} \ I_{n,\ell} \left(1, 1, \frac{mZr_{c}}{\hbar^{2}}, p_{0}\right)p_{0}^{2} \ \mathrm{d}p_{0}
 = \frac{\langle p_{0}^{m} \rangle_{n,\ell}}{\lambda^{m}}.
\end{aligned}
\end{equation}
Our desired spherically averaged Compton density can be achieved adopting the following strategy, 
\begin{equation}
J_{n,\ell} \left(\frac{\hbar^{2}}{m}, Z, r_{c}, q\right) = \int_{|q|}^{\infty} \frac{I_{n,\ell} \left(\frac{\hbar^{2}}{m}, Z, 
r_{c}, p\right)}{p} p^{2} \ \mathrm{d}p. 
\end{equation}
Here, $p=\frac{p_{0}}{\lambda}$, then $|q|=\frac{|q_{0}|}{\lambda}$ and $\mathrm{d}p = \mathrm{d}p_{0}$. Hence, applying 
Eq.~(\ref{eq:4}), 
\begin{equation}\label{eq:6}
\begin{aligned}
J_{n,\ell}\left(\frac{\hbar^{2}}{m}, Z, r_{c}, q\right) & = 
\lambda  \int_{\frac{|q_{0}|}{\lambda}}^{\infty}  \frac{I_{n,\ell} \left(1, 1, \frac{mZr_{c}}{\hbar^{2}}, p_{0}\right)}
{p_{0}}p_{0}^{2} \ \mathrm{d}p_{0} \\
& = \lambda \ J_{n,\ell} \left(1, 1, \frac{mZr_{c}}{\hbar^{2}}, q_{0}\right). 
\end{aligned}
\end{equation}
The above equation finally corresponds to the CP for CHA isoelectronic systems. Now, let us try to evaluate various momentum moments
as below, 
\begin{equation}\label{eq:7}
\begin{aligned}
\langle p^{m} \rangle_{n,\ell} & = 2(m+1) \int_{0}^{\infty} q^{m} \ J_{n,\ell}\left(\frac{\hbar^{2}}{m}, Z, r_{c}, q\right) 
\mathrm{d}q \\
  & = 2(m+1) \int_{0}^{\infty} \left(\frac{q_{0}}{\lambda}\right)^{m} \lambda 
J_{n,\ell} \left(1, 1, \frac{mZr_{c}}{\hbar^{2}}, q_{0}\right) \frac{\mathrm{d}q_{0}}{\lambda} \\
& =  \frac{1}{\lambda^{m}} \ 2(m+1) \int_{0}^{\infty} q_{0}^{m} J_{n,\ell} \left(1, 1, \frac{mZr_{c}}{\hbar^{2}}, q_{0}\right) 
\mathrm{d}q_{0}  
 = \frac{\langle p_{0}^{m}\rangle_{n,\ell}}{\lambda^{m}}. 
\end{aligned}
\end{equation} 
Thus Eqs.~(\ref{eq:5}) and (\ref{eq:7}) have provided exactly identical expression, implying that our scaling formula in 
constructing CP for isoelectronic series is correct. Following a similar argument, one can easily write, $\left\langle \frac{1}{p} 
\right\rangle_{n,\ell} = \lambda \left\langle \frac{1}{p_{0}} \right\rangle_{n,\ell}$.

Now, using Eq.~(\ref{eq:14}) and Eq.~(\ref{eq:6}) one can get,
\begin{equation}
\begin{aligned}
S^{c}_{n,\ell}\left(\frac{\hbar^{2}}{m}, Z, r_{c}, q\right)  & =  -\int J_{n,\ell}\left(\frac{\hbar^{2}}{m}, Z, r_{c}, q\right)
\ \ln J_{n,\ell}\left(\frac{\hbar^{2}}{m}, Z, r_{c}, q\right) \ \mathrm{d}q \\
& =  -\int \lambda J_{n,\ell} \left(1, 1, \frac{mZr_{c}}{\hbar^{2}}, q_{0}\right) \left(\ln \lambda +\ln J_{n,\ell} 
\left(1, 1, \frac{mZr_{c}}{\hbar^{2}}, q_{0}\right)\right) \frac{\mathrm{d}q_{0}}{\lambda}. 
\end{aligned}
\end{equation}    
Since $\int J_{n,\ell}\left(\frac{\hbar^{2}}{m}, Z, r_{c}, q\right) \mathrm{d}q = 
\int J_{n,\ell} \left(1, 1, \frac{mZr_{c}}{\hbar^{2}}, q_{0}\right) \mathrm{d}q_{0}=\frac{1}{2}$, we can write,
\begin{equation}\label{eq:8}
\begin{aligned}
S^{c}_{n,\ell}\left(\frac{\hbar^{2}}{m}, Z, r_{c}, q\right) & = -\frac{1}{2}\ln \lambda -\int J_{n,\ell} \left(1, 1, \frac{mZr_{c}}
{\hbar^{2}}, q_{0}\right) \ln J_{n,\ell} 
\left(1, 1, \frac{mZr_{c}}{\hbar^{2}}, q_{0}\right)\mathrm{d}q_{0} \\ 
S^{c}_{n,\ell}\left(\frac{\hbar^{2}}{m}, Z, r_{c}, q\right) & = -\frac{1}{2}\ln \lambda + S_{c}^{n,\ell} \left(1, 1,\frac{mZr_{c}}
{\hbar^{2}}, q_{0}\right). 
\end{aligned}
\end{equation}  
An analogous mathematical exercise will lead us to the following expression for Onicescu energy, 
\begin{equation}\label{eq:9}
E^{c}_{n,\ell}\left(\frac{\hbar^{2}}{m}, Z, r_{c}, q\right)  = \lambda E_{c}^{n,\ell} \left(1, 1,\frac{mZr_{c}}{\hbar^{2}}, 
q_{0}\right).
\end{equation}
As mentioned earlier, the present calculations are done by engaging two different wave functions, namely, (i) exact wave function 
given in Eq.~(\ref{eq:1}) and (ii) those obtained numerically through the GPS scheme. In either case, Compton profiles are 
established numerically in $p$ space. The two results are found to compliment each other in all occasions. Henceforth, we will 
report results produced from the exact wave function only.
 
\section{result and discussion}
The results will be discussed in three subsections. At first we present the CP and Compton information quantities on ground and 
some low-lying excited states of FHA. This offers an opportunity to derive the CPs in nodeless states in closed analytical form.  
This will help us to distinguish the manifestation of CP and all other measures from FHA to CHA, which is taken up in the second 
stage. Thus we will discuss and characterize the freshly built CPs of H-atom under high pressure environment, in detail. Finally, 
the above analysis will be extended to isoelectronic series of CHA.  
   
\subsection{Free H-like atom} 
In case of FHA like systems ($r_c \rightarrow \infty$) the first-order hypergeometric function modifies to associated Laguerre 
polynomial with $\mathcal{E}_{n}=-\frac{Z}{2n^{2}}$. Therefore, Eq.~(\ref{eq:1}) reduces to, 
\begin{equation}
\psi_{n,\ell}(r)= \frac{2}{n^2}\left[\frac{(n-\ell-1)!}{(n+\ell)!}\right]^{\frac{1}{2}}\left[\frac{2Z}{n}r\right]^{\ell} 
e^{-\frac{Z}{n}r} \ L_{(n-\ell-1)}^{(2\ell+1)} \left(\frac{2Z}{n}r\right). 
\end{equation}
It is well-known that the Schr\"odinger equation of H-like atom in $p$ space can be solved exactly. The \emph{exact} $p$-space wave 
function \cite{sanudo08a}, in this case, assumes the following form,
\begin{equation}
\psi_{n,\ell}(p)=\frac{n^{2}}{Z^{\frac{3}{2}}}\left[\frac{2}{\pi}\frac{(n-\ell-1)!}{(n+\ell)!}\right]^\frac{1}{2} 2^{(2\ell+2)} 
\ell! \ 
\frac{n^\ell}{ \{[\frac{np}{Z}]^2+1 \}^{\ell+2}} \left(\frac{p}{Z}\right)^\ell
C_{n-\ell-1}^{\ell+1} \left(\frac{[\frac{np}{Z}]^2-1}{[\frac{np}{Z}]^2+1}\right), 
\end{equation}
where $C_{\zeta}^{\eta}(t)$ signifies the Gegenbauer polynomial. In what follows, we attempt to provide an accurate analytical form 
of CP for nodeless states, by designing a generalized equation. From this, simplifying expressions are given for first five circular
{\color{blue}or nodeless} states. For non-circular states, they become somehow involved and we have not pursued as it is aside the 
main objective of this work. On the other hand, the exact forms of $S^{c}_{n,\ell}$ and $E^{c}_{n,\ell}$ could be derived only for 
the ground state; for all excited states, recourse has been taken to numerical method. 

We note that, in a circular state $(n-\ell)=1$ and $C_{n-\ell-1}^{\ell+1} \left(\frac{[\frac{np}{Z}]^2-1}{[\frac{np}{Z}]^2+1}\right)$ 
reduces to unity. Thus, the radial component in $p$ space simplifies to,  
\begin{equation}
\psi_{n-\ell=1}(p)  = \frac{n^{2}}{Z^{\frac{3}{2}}} \left[\frac{2}{\pi}\frac{1}{(n+\ell)!}\right]^\frac{1}{2} 2^{(2\ell+2)} \ \ell! \ 
\frac{1}{ \{[\frac{np}{Z}]^2+1 \}^{\ell+2}} \left(\frac{np}{Z}\right)^\ell. 
\end{equation}
Therefore, the spherically averaged EMD or radial density in $p$ space takes the form, 
\begin{equation}\label{eq:17}
I_{n-\ell=1}(p)  = \frac{n^{4}}{Z^{3}} \left[\frac{2}{\pi}\frac{1}{(n+\ell)!}\right] 2^{(4\ell+4)} \ (\ell!)^{2} \ 
\frac{1}{ \{[\frac{np}{Z}]^2+1 \}^{2\ell+4}} \left(\frac{np}{Z}\right)^{2\ell}.
\end{equation}  
Now, using the definition in Eq.~(\ref{eq:15}) and employing Eq.~(\ref{eq:17}), the CP can be expressed as,  
\begin{equation}\label{eq:18}
J_{n-\ell=1}(q) = \frac{1}{2}\frac{n^{4}}{Z^{3}} \left[\frac{2}{\pi} \frac{1}{(n+\ell)!}\right] 2^{(4\ell+4)} (\ell !)^{2} 
\int_{|q|}^{\infty} \frac{1}{ \{[\frac{np}{Z}]^2+1 \}^{2\ell+4}} \left(\frac{np}{Z}\right)^{2\ell} p \ \mathrm{d}p.
\end{equation}      
By setting $\left(\frac{np}{Z}\right)^{2}=y$, Eq.~(\ref{eq:18}) can be rewritten as, 
\begin{equation}\label{eq:19} 
J_{n-\ell=1}(q)= \frac{n^{2}}{4Z}\left[\frac{2}{\pi} \frac{1}{(n+\ell)!}\right] 2^{(4\ell+4)} (\ell !)^{2} 
\int_{\left(\frac{nq}{Z}\right)^{2}}^{\infty} \frac{y^{\ell}}{(y+1)^{2\ell+4}}\ \mathrm{d}y.
\end{equation}
Now adopting the binomial expression and replacing $y^{\ell}=(y+1)^{\ell}-\sum_{k=1}^{\ell}\frac{\ell!}{k!(\ell-k)!}\ y^{(\ell-k)}$,  
in Eq.~(\ref{eq:19}), one arrives at the following generalized form,
\begin{equation}\label{eq:20} 
J_{n-\ell=1}(q)= \frac{n^{2}}{4Z}\left[\frac{2}{\pi} \frac{1}{(n+\ell)!}\right] 2^{(4\ell+4)} (\ell !)^{2} 
\int_{\left(\frac{nq}{Z}\right)^{2}}^{\infty} \frac{[(y+1)^{\ell}-\sum_{k=1}^{\ell}\frac{\ell!}{k!(\ell-k)!}\ y^{(\ell-k)}]}
{(y+1)^{2\ell+4}} \ \mathrm{d}y.
\end{equation}
This is our general expression for $J_{n-l}(q)$ in circular states of H-like atoms. Using appropriate values of $n, \ell$ in 
Eq.~(\ref{eq:20}), following expressions can be written for five such lowest states:
\begin{eqnarray}
J_{1s}(q) & = & \frac{8}{3\pi Z} \frac{1}{\left[\left(\frac{q}{Z}\right)^{2}+1\right]^{3}},\label{eq:21} \\
J_{2p}(q) & = & \frac{64}{15\pi Z} \frac{\left[5 \left(\frac{2q}{Z}\right)^{2}+1\right]}
{\left[\left(\frac{2q}{Z}\right)^{2}+1\right]^{5}},\label{eq:22} \\
J_{3d}(q) & = & \frac{3072}{525\pi Z} \frac{\left[21 \left(\frac{3q}{Z}\right)^{4}+7 \left(\frac{3q}{Z}\right)^{2}+1\right]}
{\left[\left(\frac{3q}{Z}\right)^{2}+1\right]^{7}},\label{eq:23} \\
J_{4f}(q) & = & \frac{16384}{2205\pi Z} \frac{\left[84 \left(\frac{4q}{Z}\right)^{6}+ 36 \left(\frac{4q}{Z}\right)^{4}+9 
\left(\frac{4q}{Z}\right)^{2}+1\right]}
{\left[\left(\frac{4q}{Z}\right)^{2}+1\right]^{9}},\label{eq:24} \\
J_{5g}(q) & = & \frac{131072}{14553\pi Z} \frac{\left[330\left(\frac{5q}{Z}\right)^{8} + 165 \left(\frac{5q}{Z}\right)^{6}+ 55 
\left(\frac{5q}{Z}\right)^{4}+ 
11 \left(\frac{5q}{Z}\right)^{2}+1\right]}{\left[\left(\frac{5q}{Z}\right)^{2}+1\right]^{11}}\label{eq:25}.
\end{eqnarray}
It is interesting to mention that the expressions for $1s, 2p$ states in Eq.~(\ref{eq:21}) and (\ref{eq:22}) are the same
as those given long times ago in \cite{duncanson45}. However the generalized expression, Eq.~(\ref{eq:20}) apparently has not 
been reported before. {\color{blue} Similar exercise can also be undertaken for non-circular states as well. But these are 
cumbersome and outside the main focus of this work, as they have no bearing on the general conclusions made here. 
That is why they are not pursued here.} Now following Eq.(\ref{eq:16}), one can easily write,
\begin{equation}\label{eq:50} 
\begin{aligned}
	\left\langle \frac{1}{2p} \right\rangle_{1s} & =   \frac{8}{3\pi Z}=\frac{0.84882636}{Z}, \ \ \ \ \ \ \ \ \ \  
	\left\langle \frac{1}{2p} \right\rangle_{2p}   =   \frac{64}{15\pi Z}=\frac{1.35812218}{Z}, \\
	\left\langle \frac{1}{2p} \right\rangle_{3d} & =   \frac{3072}{525\pi Z}=\frac{1.86256756}{Z}, \ \ \ \ \ \ \  
	\left\langle \frac{1}{2p} \right\rangle_{4f}   =   \frac{16384}{2205\pi Z}=\frac{2.36516515}{Z}, \\
	\left\langle \frac{1}{2p} \right\rangle_{5g} & =   \frac{131072}{14553\pi Z}=\frac{2.866866859}{Z}.
\end{aligned}
\end{equation} 
{\color{blue}Now the focus is to investigate $E^{c}_{n,\ell}$ and $S^{c}_{n,\ell}$ of FHA. The ground-state 
$E^{c}_{1,0}$ and $S^{c}_{1,0}$ possess following closed forms (details are provided in Appendix~I $\&$ II respectively). 

\begin{equation} \label{eq:30a}
E^{c}_{1,0} = \omega^{2}  = \left(\frac{8}{3\pi}\right)^{2} \frac{1}{Z} B\left(\frac{11}{2},\frac{11}{2}\right) 
= \left(\frac{8}{3\pi}\right)^{2} \frac{1}{Z} \ \frac{\Gamma (\frac{11}{2}) \Gamma (\frac{11}{2})}{\Gamma (11)} = 
\left[\frac{7}{8\pi Z}\right]
\end{equation}

\begin{equation}\label{eq:33a}
S^{c}_{1,0}=\frac{1}{2} \ln 24 \pi + \frac{1}{2} \ln Z- \frac{7}{4}.
\end{equation}}

Finally, when $Z=1$, Eqs.~(\ref{eq:30a}), (\ref{eq:33a}) produce $E^{c}_{1s}=\frac{7}{8\pi}=0.27852115$ and $S^{c}_{1s}= \frac{1}{2} 
\ln 24\pi-\frac{7}{4}=0.411391858$. Here, we have restricted ourselves to $S^{c}_{n,\ell}, E^{c}_{n,\ell}$ in $1s$ state only.
{\color{blue}But for higher states, it is difficult to derive their closed forms. However, numerical values can be achieved.
Results for some of these states are provided in Table~II of Sec.~III.B.} These outcomes in $1s$ provide us an idea to predict 
the general trend in $S^{c}_{n,l}, E^{c}_{n,l}$ with respect to $Z$. Later, Sec.~III.C establishes that $S^{c}_{n,l}$ linearly 
increases with $Z$, whereas, $E^{c}_{n,l}$ is inversely proportional to $Z$.
   
\begin{figure}                         
\begin{minipage}[c]{0.32\textwidth}\centering
\includegraphics[scale=0.56]{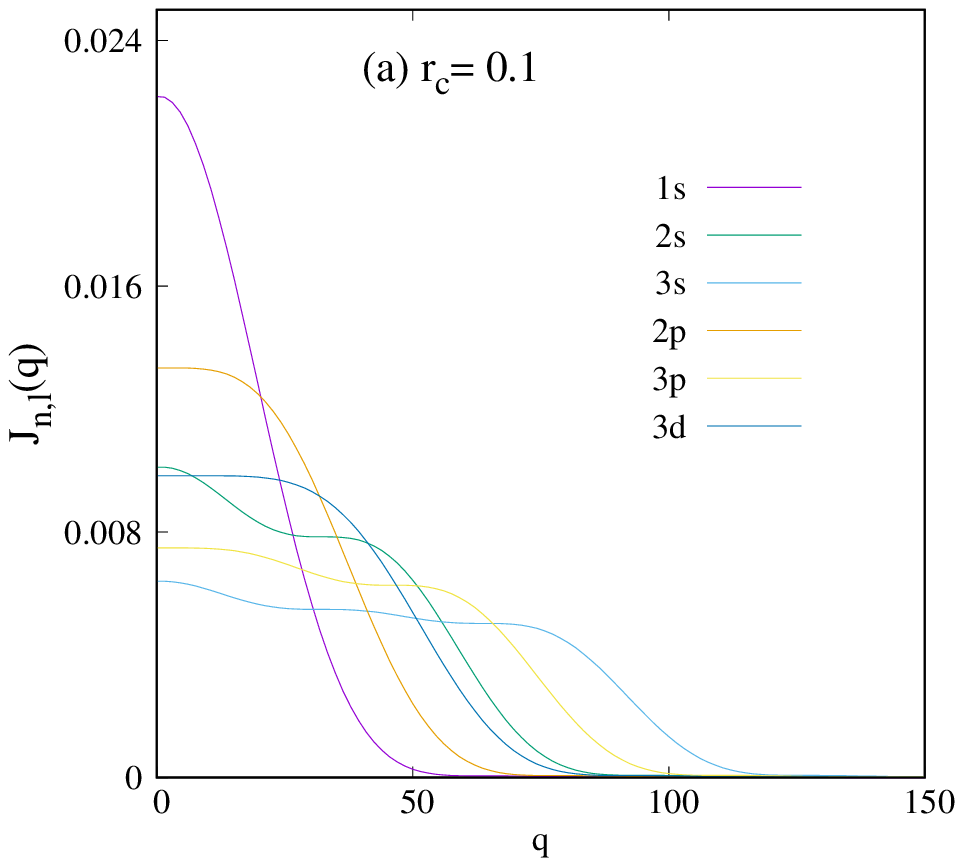}
\end{minipage}%
\vspace{1mm}
\begin{minipage}[c]{0.32\textwidth}\centering
\includegraphics[scale=0.56]{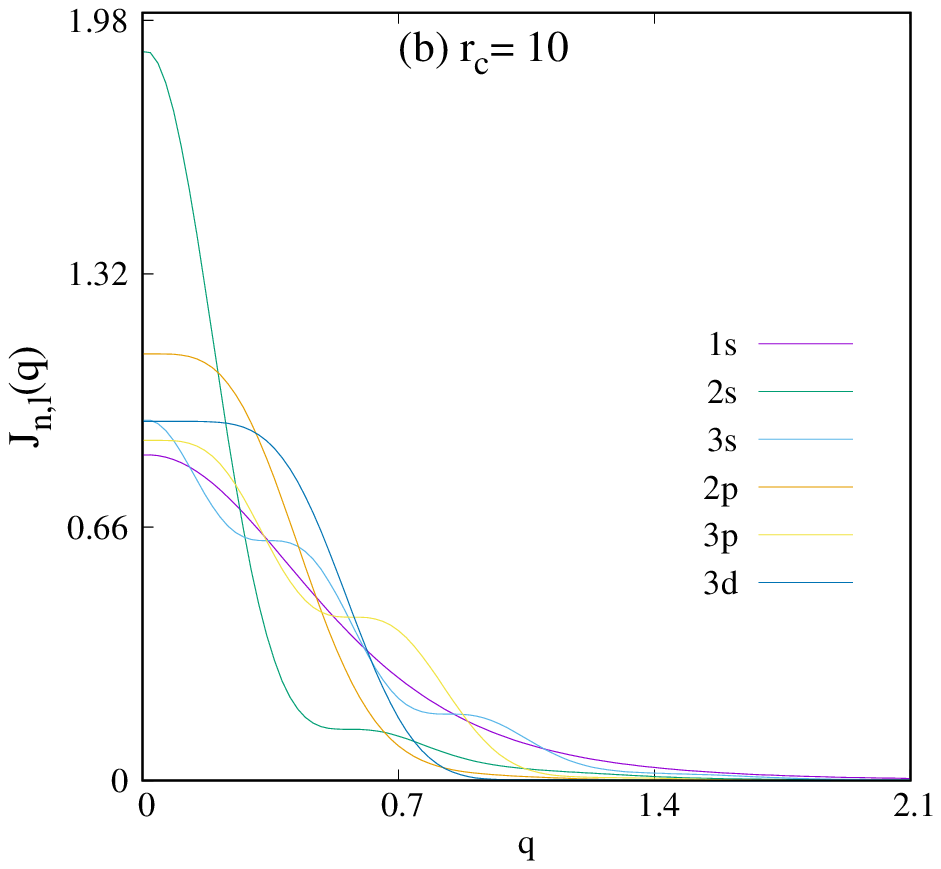}
\end{minipage}%
\vspace{1mm}
\begin{minipage}[c]{0.32\textwidth}\centering
\includegraphics[scale=0.56]{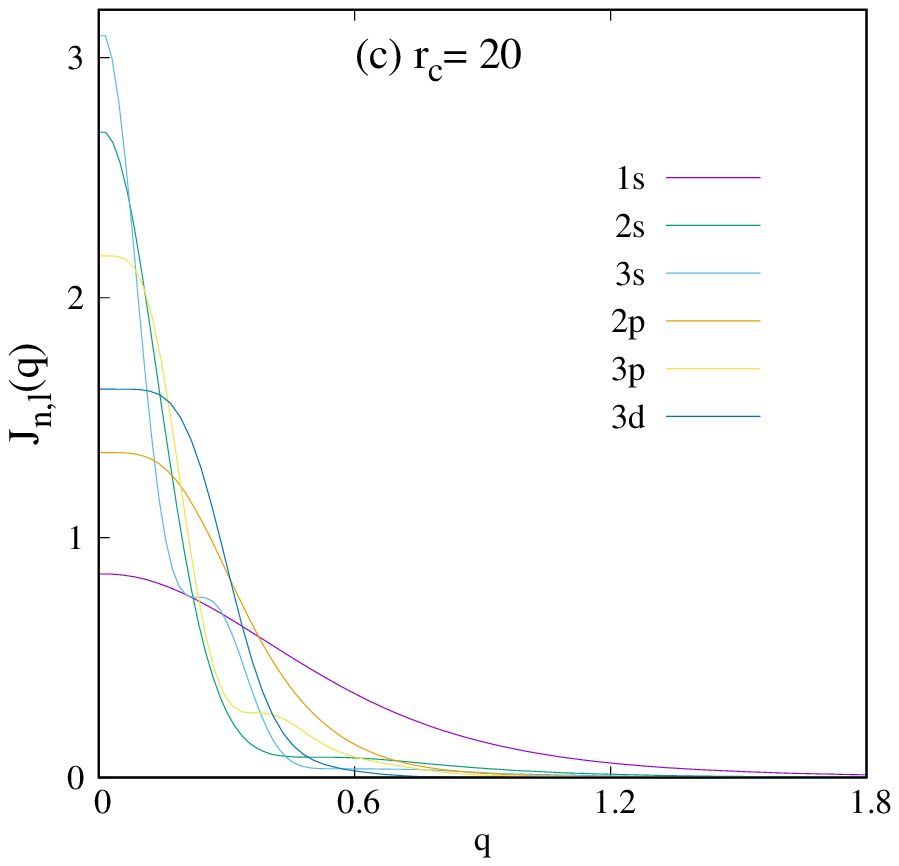}
\end{minipage}%
\caption{CP for all six $n \leq 3$ states in CHA at $r_c$ values $0.1, 10, 20$ in panels (a)-(c). See text for details.}
\end{figure}

\subsection{Confined H atom}
Now we shift our focus to the central theme of this work, \emph{viz.},CHA. The numerically calculated CPs, obtained from 
Eq.~(\ref{eq:15}), are depicted in Fig.~1. The six lowest states
(1s, 2s, 3s, 2p, 3p, 3d) having principal quantum number $n \leq 3$ at three $r_{c}$ values 0.1, 10, 20, identifying strong, 
intermediate and weak confinement regions are offered in panels (a)-(c) respectively. Accuracy of these CPs have been thoroughly 
verified by independently computing selected expectation values, $\left\langle \frac{1}{2p} \right\rangle, \langle p^{2} 
\rangle$ and $\langle p^{4}\rangle$ in accordance with the right side of Eq.~(\ref{eq:16}). {\color{blue} The first one is 
directly connected to Compton profile, whereas $\langle p^{2} \rangle$ is proportional to kinetic energy. Similarly,     
$\langle p^{4} \rangle$ is related to relativistic kinetic energy. The results obtained using Eq.~(\ref{eq:16})} are collected in 
Table~I, for all six states at seven different box radii, namely, $0.1, 0.2, 0.5, 1, 5, 10, \infty$. {\color{blue} It is a 
known fact that, these momentum moments can also be computed by either (a) using eigenfunction (Eq.~(\ref{eq:1})) in $r$ space or 
(b) employing EMD in $p$ space. The results obtained from Eq.~(\ref{eq:16}) are in complete agreement with the reference values 
achieved for both free and confined H atom by adopting the above two methods. Due to the lack of space, only $1s$ state
results of CHA are reported in the footnote.} All the quantities in this table match up to all 
the decimal points reported. {\color{blue} This convergence acts as a probe about the accuracy of the constructed CPs.}    

\begingroup           
\squeezetable
\begin{table}
\caption{$\langle \frac{1}{2p}\rangle, \langle p^{2} \rangle, \langle p^{4} \rangle$ for six low-lying states of CHA at 
seven $r_{c}$. See text for detail.}
\centering
\begin{ruledtabular}
\begin{tabular}{l|l|lllllll}
State & Property  &  $r_c=0.1$ & $r_c=0.2$ & $r_c=0.5$ & $r_c=1$ & $r_c=5$ & $r_c=10$ & $r_c=\infty$   \\            
\hline
& $\langle \frac{1}{2 p}\rangle$ &
0.0221757 & 0.044248 & 0.109759 & 0.215997 & 0.77876 & 0.84811 & 0.848826$^{\S}$  \\
1s & $\langle p^{2} \rangle^{\dagger}$  & 
987.184273 & 246.969146 & 39.724028 & 10.146273 & 1.049068 & 1.000024 & 1.000000 \\
& $\langle p^{4} \rangle^{\ddag}$ &
975768.196166 & 61315.262248 & 1636.153662 & 120.909615 & 5.170882 & 4.999913 &  4.99999 \\           
\hline
& $\langle \frac{1}{2 p}\rangle$ &
0.010106 & 0.020298 & 0.051411 & 0.105489 & 0.76495 & 1.89844 & 2.716244 \\
2s & $\langle p^{2} \rangle$  & 
3947.9553 & 987.07315 & 158.02331 & 39.580724 & 1.5360322 & 0.3720624 & 0.249999  \\
& $\langle p^{4} \rangle$ &
15590054.13617 & 975260.28106 & 25133.40985 & 1611.86976 & 5.62197 & 1.14715 & 0.812499 \\
\hline
& $\langle \frac{1}{2 p}\rangle$ &  
0.00639 & 0.01278 & 0.03203 & 0.06439 & 0.35856 & 0.93869 & 4.87468 \\
3s & $\langle p^{2} \rangle$  & 
8882.714743 & 2220.730969 &  355.373136 & 88.888824 & 3.552168 & 0.829053 & 0.111111 \\
& $\langle p^{4} \rangle$ &
78909217.8106 & 4933319.9437 &  126570.0272 & 7976.4667 & 17.372167 & 2.127063 & 0.259259 \\
\hline
& $\langle \frac{1}{2 p}\rangle$ &
0.013336 & 0.026655 & 0.066504 & 0.132535 & 0.6330957 & 1.111514 & 1.358122$^{\S}$   \\
2p & $\langle p^{2} \rangle$  & 
2019.114160 & 504.809879 &  80.805690 &  20.235456 & 0.873647 & 0.313022 & 0.249999     \\
& $\langle p^{4} \rangle$ &
4077013.91352 & 254881.58402 &   6537.62441 &    411.62171 &    0.91591 &    0.185835 &    0.145833 \\
\hline
& $\langle \frac{1}{2 p}\rangle$ &
0.007478 & 0.014966 & 0.037522 & 0.075258 & 0.393877 & 0.886223 & 2.910261  \\
3p & $\langle p^{2} \rangle$  & 
5967.983470 & 1492.019813 &  238.750084 &  59.711670 & 2.417646 & 0.614379 & 0.111111  \\
& $\langle p^{4} \rangle$ &
35617508.568154 & 2226295.030051 & 57029.814217 & 3572.844244 & 6.264438 & 0.535440 & 0.061728  \\
\hline
& $\langle \frac{1}{2 p}\rangle$ &
0.009830 & 0.019656 & 0.049111 & 0.098056 & 0.482851 & 0.935688 & 1.862567$^{\S}$   \\
3d & $\langle p^{2} \rangle$  & 
3321.761311 & 30.451738 &  132.885252 &  33.233236 & 1.348000 & 0.357810 & 0.111111  \\
& $\langle p^{4} \rangle$ &
11034178.1130 & 689670.182588 & 17661.762592 & 1105.290838 & 1.860854 & 0.144078 & 0.022222  \\
\end{tabular}
\end{ruledtabular}
\begin{tabbing} 
{\color{blue}$^{\dagger}$ Near-exact values of $\langle p^{2} \rangle$ for $1s$ state in CHA obtained by using Eq.~(\ref{eq:1}) at 
$r_c=0.1, 0.2, 0.5, 1, 5, 10, \infty$ are: $987.18427303, 246.96914695,$}\\
{\color{blue}$39.724028075, 10.146272936,  1.0490688068,  1.0000248591,  1$ respectively.} \\
{\color{blue}$^{\ddag}$ Near-exact values of $\langle p^{4} \rangle$ for $1s$ state in CHA obtained by using Eq.~(\ref{eq:1}) at 
$r_c=0.1, 0.2, 0.5, 1, 5, 10, \infty$ are: $975768.19616638, 61315.26224861,$} \\
{\color{blue}$1636.15366289, 120.90961564, 5.17088297, 4.99991343,  5$ respectively.} \\
{\color{blue}$^{\S}$Exact values in FHA, from Eq.~(\ref{eq:50}), for 1s, 2p, 3d states are: $0.84882636, 1.35812218, 1.86256756$ respectively.}   
\end{tabbing}
\end{table} 

An in-depth analysis of panels (a)-(c) of Fig.~1 suggests that, there appears multiple humps in a given CP, which become 
prominent with rise in $r_{c}$. Interestingly, in a particular state, the number of such humps corroborate the number of radial 
nodes in that state. Additionally, the sharpness of these CPs enhances with weakening of pressure effect. Similarly a careful
examination of Table~I reveals that, for a given state characterized by quantum numbers $(n,\ell)$, the initial intensity 
$\left\langle \frac{1}{2p} \right\rangle_{n,\ell} = J_{n,\ell}(q=0)$ tends to grow with $r_c$, while both $\langle p^{2} 
\rangle_{n,\ell}$ and $\langle p^{4} \rangle_{n,\ell}$ decay. The dependence of $\langle p^{2} \rangle_{n,\ell}$ and 
$\langle p^{4} \rangle_{n,\ell}$ on $n,\ell$ quantum number at \emph{free} and \emph{confined} condition is very clear and 
supports our previous observation \cite{mukherjee18}. It may be recalled that, in FHA $\langle p^{2} \rangle$ 
is independent of $\ell$ and decays with rise in $n$. But in CHA it varies with both $n,\ell$; at a fixed $\ell$, it grows 
with $n$, whereas at a given $n$, decreases as $\ell$ progresses. In contrast, $\langle p^{4} \rangle$, in both CHA and 
FHA, depends on $n,\ell$, producing similar pattern as found here.  


          
\begin{figure}                         
\begin{minipage}[c]{0.5\textwidth}\centering
\includegraphics[scale=0.78]{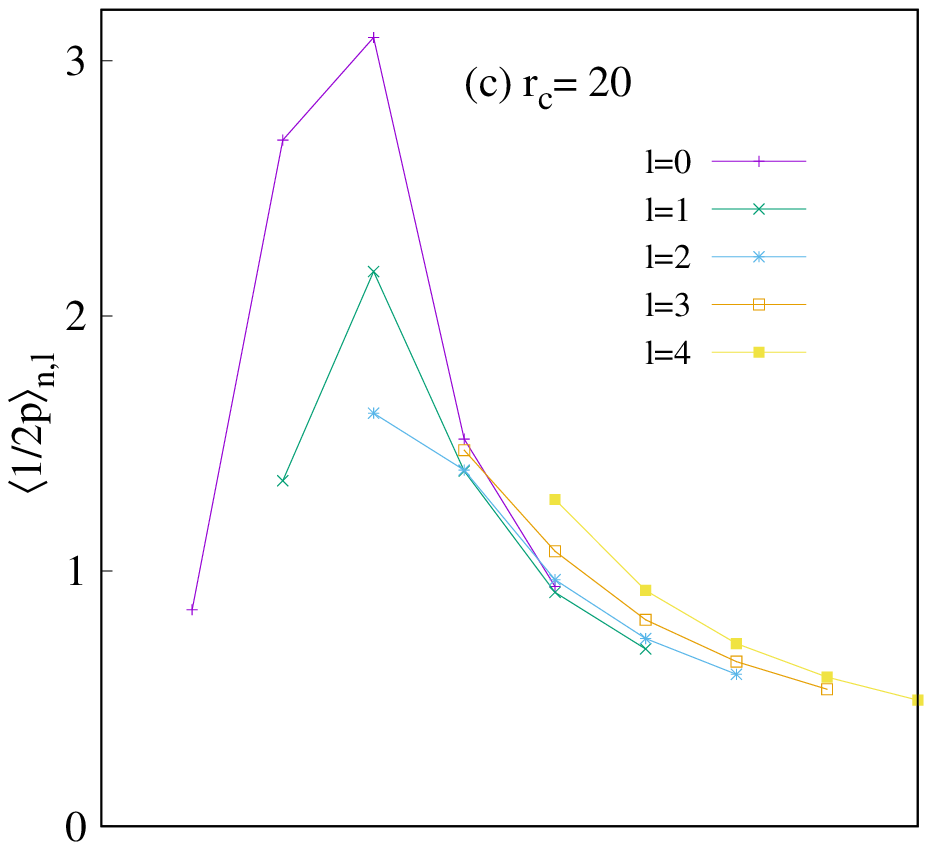}
\end{minipage}%
\begin{minipage}[c]{0.5\textwidth}\centering
\includegraphics[scale=0.78]{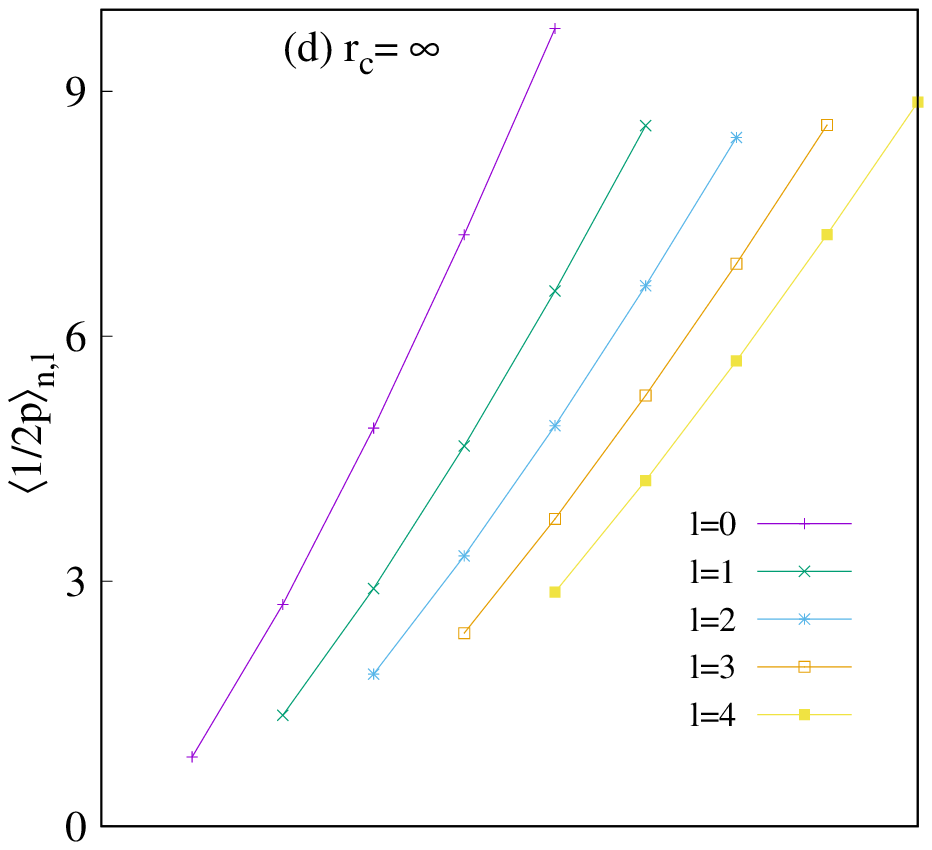}
\end{minipage}%
\vspace{2mm}
\hspace{2mm}
\begin{minipage}[c]{0.5\textwidth}\centering
\includegraphics[scale=0.82]{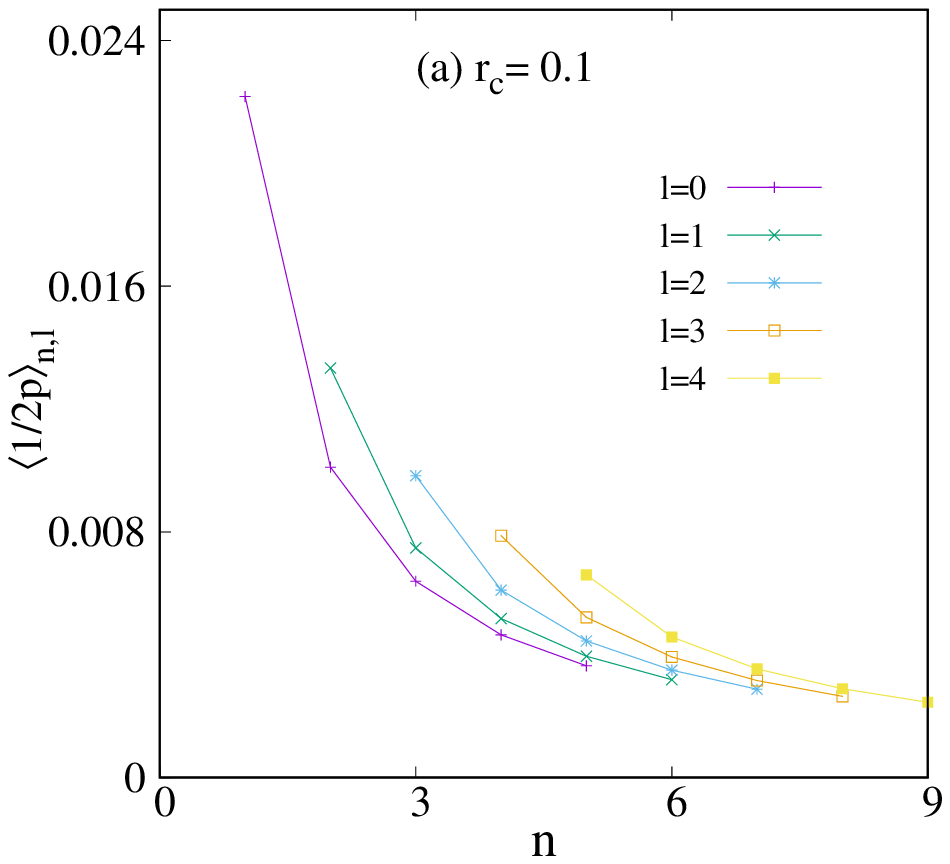}
\end{minipage}%
\begin{minipage}[c]{0.5\textwidth}\centering
\includegraphics[scale=0.82]{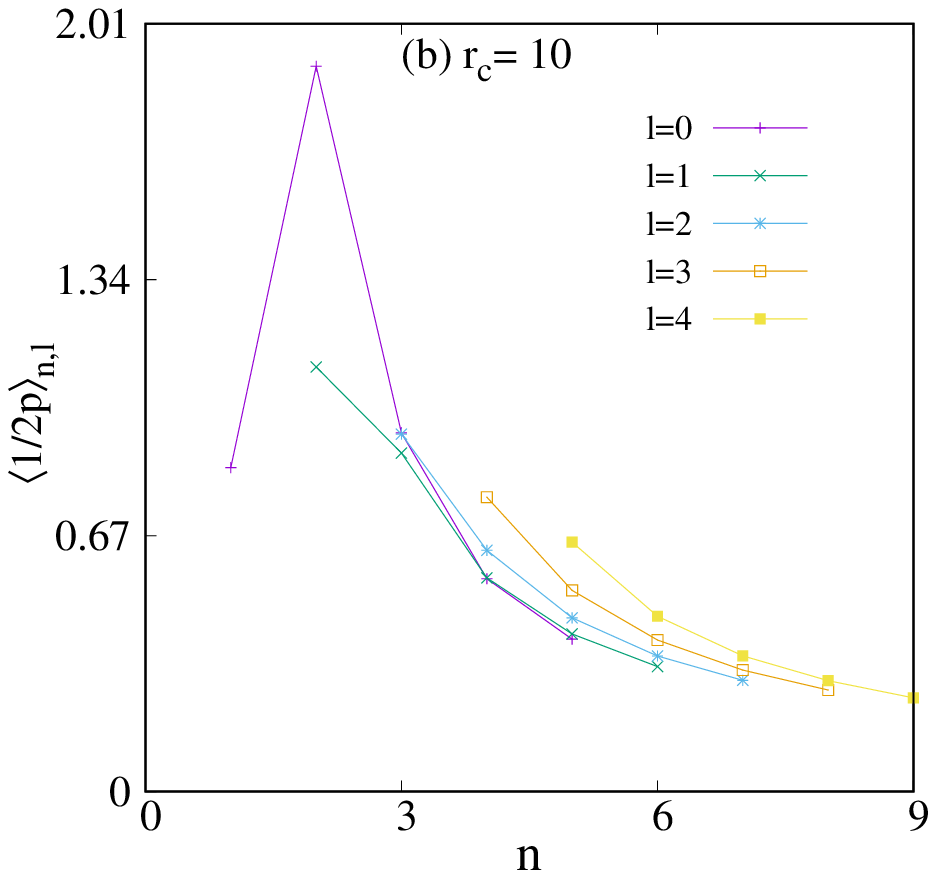}
\end{minipage}%
\caption{$\left\langle \frac{1}{2p} \right\rangle_{n,\ell}$ versus $n$ for $s,p,d,f,g$ states of CHA, at four $r_c$'s in panels 
(a)-(d). See text for details.}
\end{figure}

In order to get a further insight, dependency of $\left\langle \frac{1}{2p} \right\rangle_{n,\ell}$ on $n, \ell$ quantum numbers 
is addressed in Fig.~2. It may be noted that $\left\langle \frac{1}{2p} \right\rangle$ and $\langle p^{m} \rangle$ are inversely 
proportional to each other. Therefore, this particular study will automatically comment about the qualitative nature of 
$\langle p^{m} \rangle$ with change in $n,\ell$. From panel (a), it is seen that, at strong confinement region ($r_{c}=0.1$), 
for all $\ell$ initial intensity falls off at the end with rise in $n$. However, at a fixed $n$, it advances with rise in $l$. 
Consequently, with increase in radial nodes (so does kinetic energy), $\left\langle \frac{1}{2p} \right\rangle$ decreases 
and \emph{vice-versa}. On the contrary, as $r_{c} \rightarrow \infty$ in panel (d), a completely opposite trend of 
$\left\langle \frac{1}{2p} \right\rangle$ with respect to $n, \ell$ emerges, where increase in radial nodes leads to the 
accumulation of $\left\langle \frac{1}{2p} \right\rangle$. But as is known, in FHA, kinetic energy depends only on 
$n$ $\left(\frac{Z}{2n^2}\right)$. Moreover, panels (b) and (c) show the appearance of a maximum point, whose positions shift 
to right as $r_c$ goes up. Actually, these two segments serve as a missing link in the evolution of H atom from CHA to FHA.        

From the foregoing discussion of Figs.~1 and 2 as well as Table~I, it is quite clear that the constructed CPs presented above 
are sufficiently accurate. Several interesting features may be emphasized, \emph{viz.}, (i) enhancement of sharpness of CP's with 
relaxation of confinement (ii) alteration of behavioral pattern of momentum moments from CHA to FHA (iii) influence of radial 
nodes on CP. But the reasons behind these are not clear. Moreover, this analysis is inadequate to explain the significance of such 
broadening/sharpening of CP with pressure. In this context, information based measures like $S,E$ may serve our purpose. 
Estimation of these two quantities may provide the required opportunity for CP in CHA, which we will be pursued now.

\begingroup           
\squeezetable
\begin{table}
\caption{$S^{c}_{n,l}, E^{c}_{n,l}$ for 1s, 2s, 3s, 2p, 3p, 3d states in CHA at ten $r_{c}$'s. See text for detail.}
\centering
\begin{ruledtabular}
\begin{tabular}{l|llllll}
$r_{c}$ & \multicolumn{6}{c}{$S^{c}_{n,l}$} \\
\cline{2-7}
    & 1s &  2s & 3s & 2p & 3p & 3d    \\            
    \hline
0.1 & 2.14595 & 2.48251  & 2.67471 & 2.32217  & 2.58115  & 2.44189  \\   
0.2 & 1.79988 & 2.13588  & 2.32852 & 1.97559  & 2.23462  & 2.09531  \\   
0.5 & 1.34408 & 1.67826  & 1.87058 & 1.51781  & 1.77381  & 1.63627  \\   
1   & 1.00466 & 1.33280  & 1.52472 & 1.17187  & 1.43015  & 1.29064  \\   
2.5 & 0.59679 & 0.87982  & 1.06517 & 0.71811  & 0.97342  & 0.83338   \\   
5   & 0.42686 & 0.51831  & 0.73501 & 0.38977  & 0.63244  & 0.48999   \\   
8   & 0.41196 & 0.22351  & 0.49872 & 0.20212  & 0.40834  & 0.26330   \\   
10  & 0.41188 & 0.09081  & 0.37264 & 0.13676  & 0.30206  & 0.16081   \\   
12  & 0.41174 & 0.00438  & 0.25721 & 0.10534  & 0.21138  & 0.08282   \\   
$\infty$ & 0.411392$^{\P}$ & $-$0.078101 & $-$0.297814   & 0.082991   & $-$0.195942   &  $-$0.117328   \\
\hline
 & \multicolumn{6}{c}{$E^{c}_{n,l}$} \\
\cline{2-7}
0.1 & 0.007964 & 0.003813 & 0.002537 & 0.005390 & 0.003092 & 0.004166     \\   
0.2 & 0.015895 & 0.007625 & 0.005072 & 0.010774 & 0.006183 & 0.008331     \\   
0.5 & 0.039441 & 0.019042 & 0.012669 & 0.026892 & 0.015469 & 0.020822     \\   
1   & 0.077621 & 0.038045 & 0.025293 & 0.053608 & 0.030857 & 0.041580     \\   
2.5 & 0.178217 & 0.096167 & 0.063166 & 0.132347 & 0.076835 & 0.103511      \\   
5   & 0.265091 & 0.221328 & 0.127019 & 0.255510 & 0.152718 & 0.204978      \\   
8   & 0.278295 & 0.447362 & 0.217004 & 0.378673 & 0.244927 & 0.322178      \\   
10  & 0.278502 & 0.596075 & 0.291224 & 0.438985 & 0.311512 & 0.3961003      \\   
12  & 0.278514 & 0.713118 & 0.379233 & 0.474932 & 0.387316 & 0.465209      \\   
$\infty$ & 0.278521$^{\P}$ & 0.860646 & 1.394031 & 0.505837 & 1.020566 & 0.732503   \\
\end{tabular}
\end{ruledtabular}
\begin{tabbing} 
$^{\P}$Exact values of $S^{c}_{n,\ell}, E^{c}_{n,\ell}$ for 1s state in FHA, from Eqs.~(\ref{eq:33}) and (\ref{eq:30}) 
are: 0.411391858 and 0.27852115 respectively.   
\end{tabbing}
\end{table}

Both $S^{c}_{n,\ell}$ and $E^{c}_{n,\ell}$ values for six states (same as Table~I) of CHA, at ten different $r_{c}$'s, 
($0.1,0.2, 0.5, 1, 2.5, 5, 8, 10, 12, \infty$) are given in Table~II. In all cases, $S^{c}_{n,\ell}$ regresses and $E^{c}_{n,\ell}$ 
advances with progress in $r_c$. Since $S$ represents the measure of uncertainty in a given distribution, a sharp distribution
has less uncertainty and hence low $S$ and a broad distribution corresponds to larger uncertainty and hence greater $S$. 
Therefore, for an arbitrary-$(n,\ell)$ state, CP gets flattened with rise in pressure and conversely, with increase in cavity 
radius, CP becomes sharp. Thus, at strong confinement region, we get broader profile {\color{blue}having higher uncertainty 
in kinetic energy as well as in momentum.} Additionally at high pressure regime, boundedness of an electron enhances {\color{blue}leading to 
increment in momentum uncertainty}. Moreover, the rate of dissipation of Compton energy is slower in CHA compared to FHA. Note
that CPs relate to kinetic energy dissipation curves; sharp CPs {\color{blue}(less uncertainty in momentum)} release kinetic energy faster 
compared to a broader CP {\color{blue}(higher uncertainty momentum)}. Finally, one can conjecture that, a state having higher kinetic energy 
(implies larger $S^c$ and a resultant broad CP) releases Compton energy slowly. 
              
\begin{figure}                         
\begin{minipage}[c]{0.5\textwidth}\centering
\includegraphics[scale=0.78]{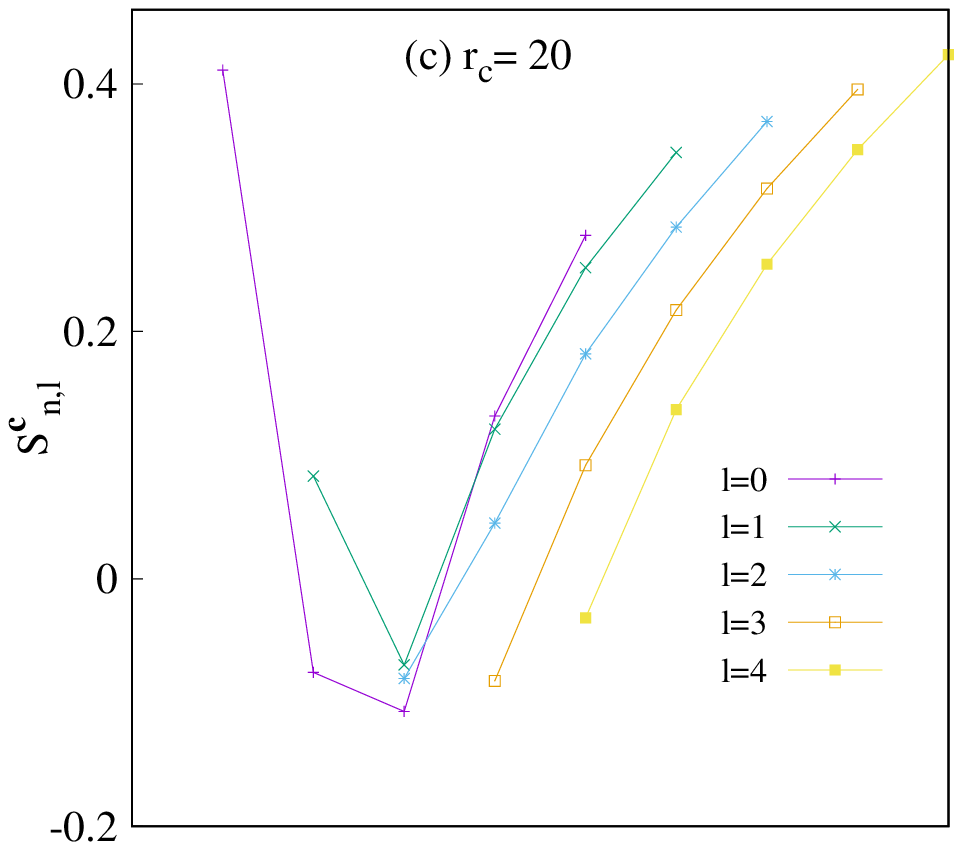}
\end{minipage}
\begin{minipage}[c]{0.5\textwidth}\centering
\includegraphics[scale=0.78]{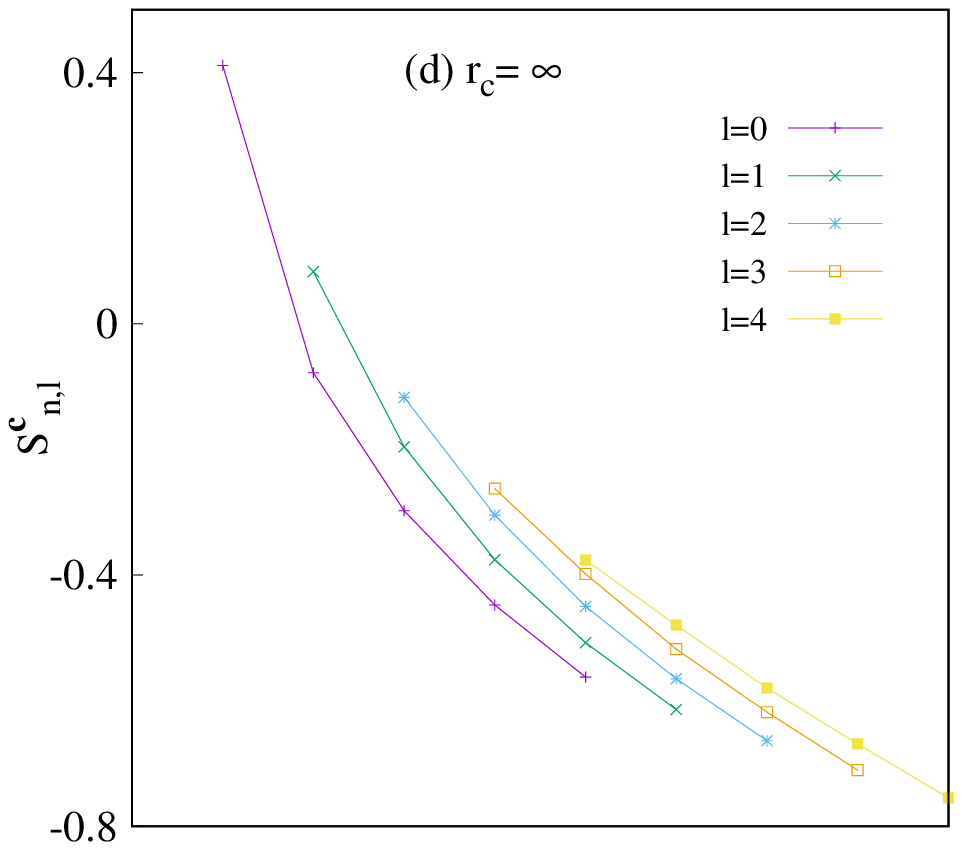}
\end{minipage}%
\vspace{2mm}
\hspace{2mm}
\begin{minipage}[c]{0.5\textwidth}\centering
\includegraphics[scale=0.82]{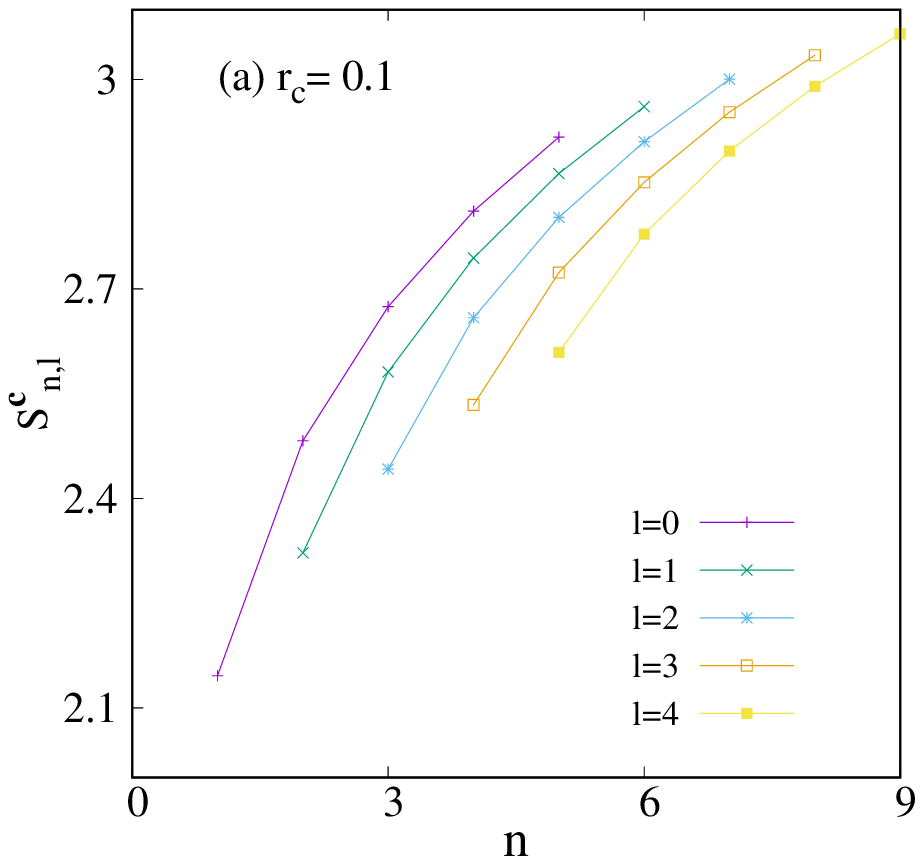}
\end{minipage}%
\begin{minipage}[c]{0.5\textwidth}\centering
\includegraphics[scale=0.82]{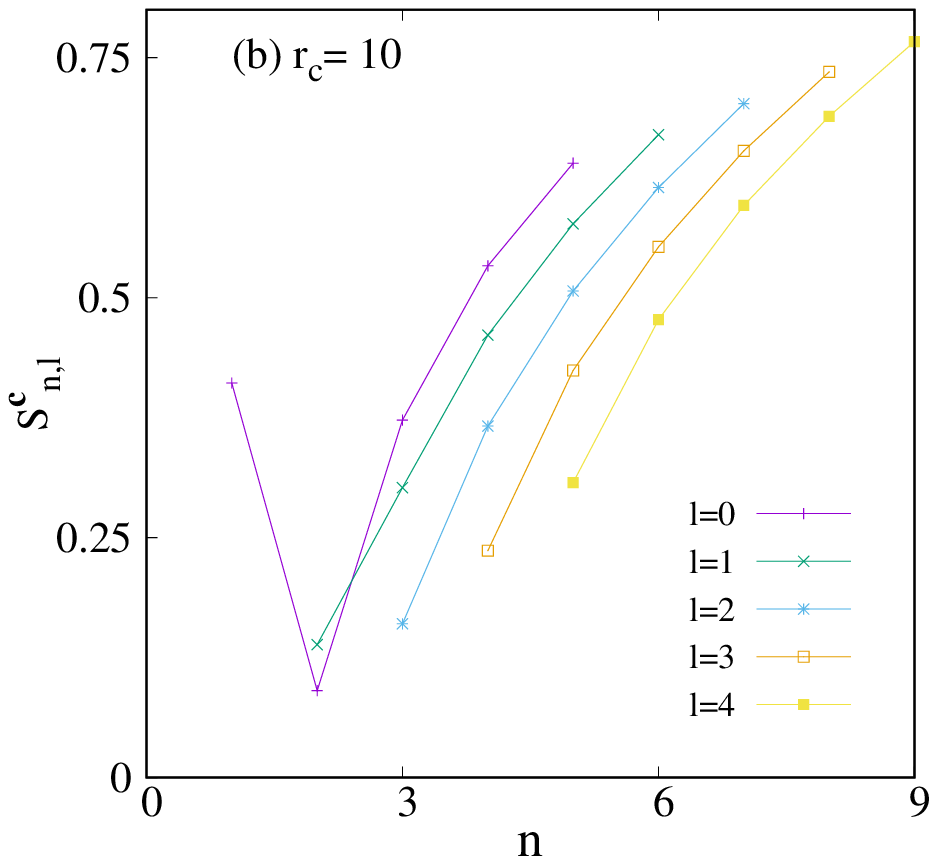}
\end{minipage}%
\caption{$S^{c}_{n,\ell}$ versus $n$ for $s,p,d,f,g$ states of CHA, at four $r_c$'s in panels 
(a)-(d). See text for details.}
\end{figure}

\begin{figure}                         
\begin{minipage}[c]{0.5\textwidth}\centering
\includegraphics[scale=0.78]{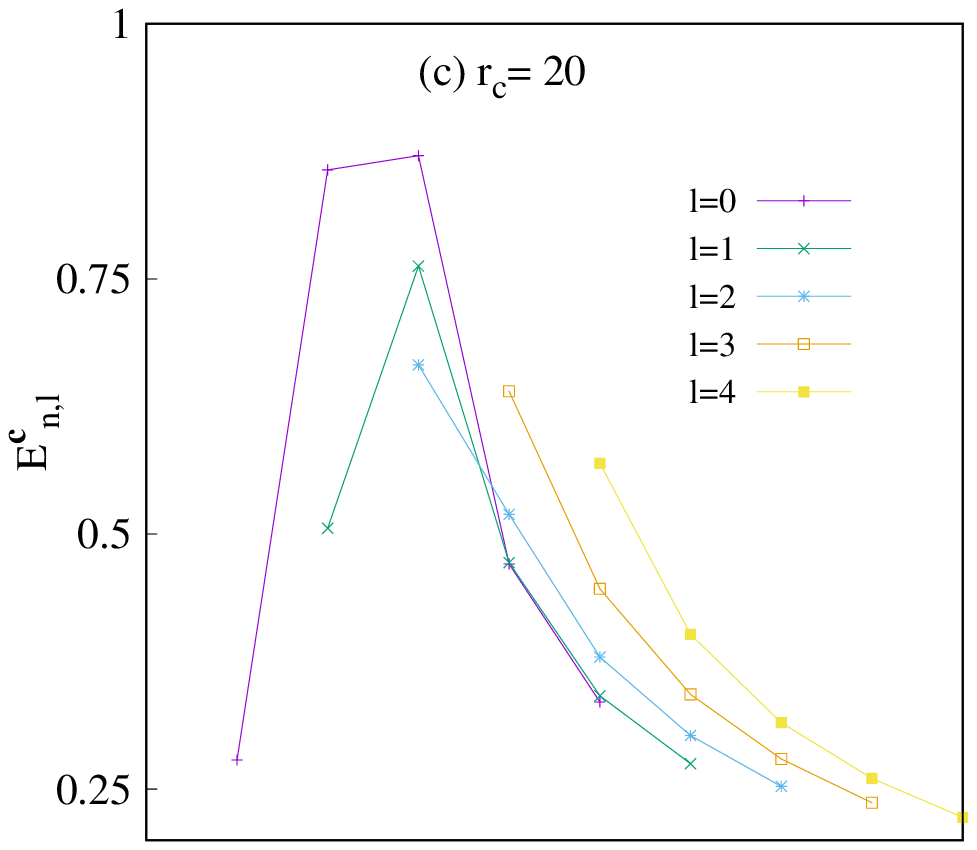}
\end{minipage}%
\begin{minipage}[c]{0.5\textwidth}\centering
\includegraphics[scale=0.78]{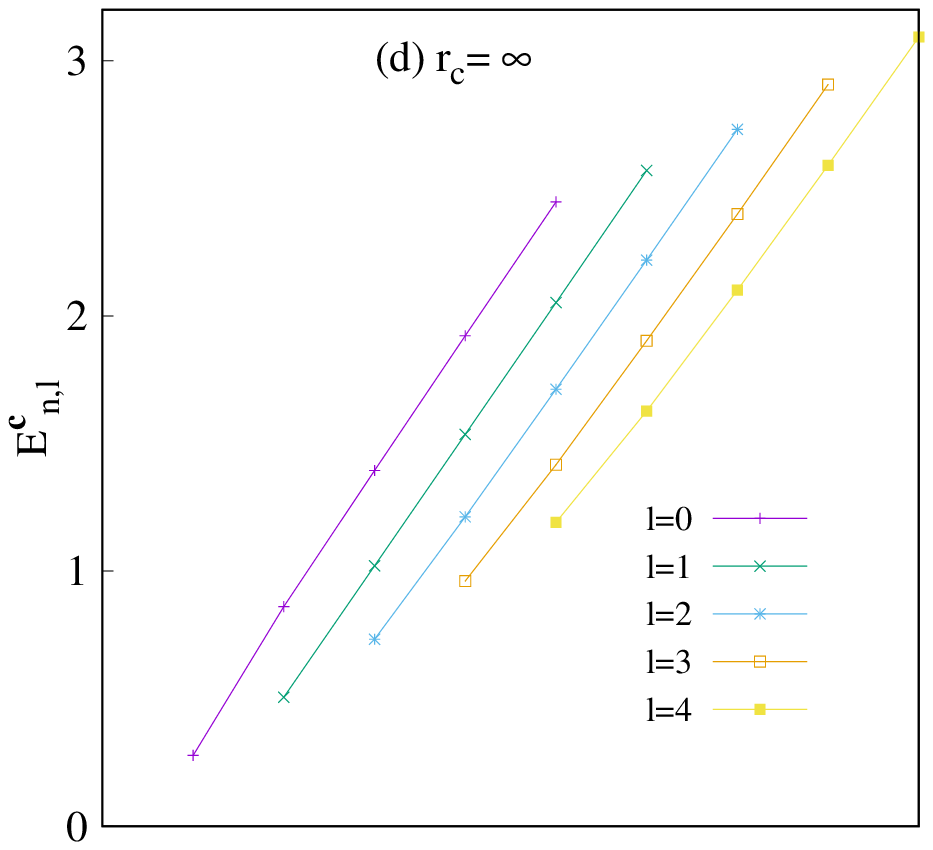}
\end{minipage}%
\vspace{2mm}
\hspace{2mm}
\begin{minipage}[c]{0.5\textwidth}\centering
\includegraphics[scale=0.82]{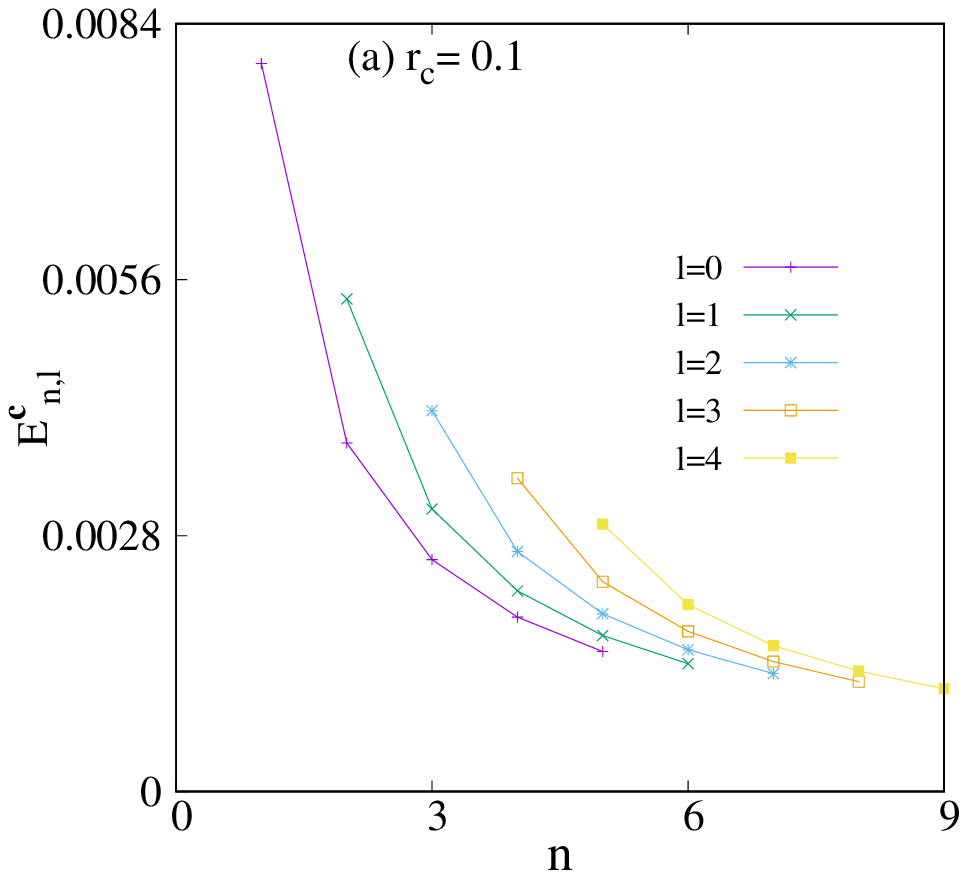}
\end{minipage}%
\begin{minipage}[c]{0.5\textwidth}\centering
\includegraphics[scale=0.82]{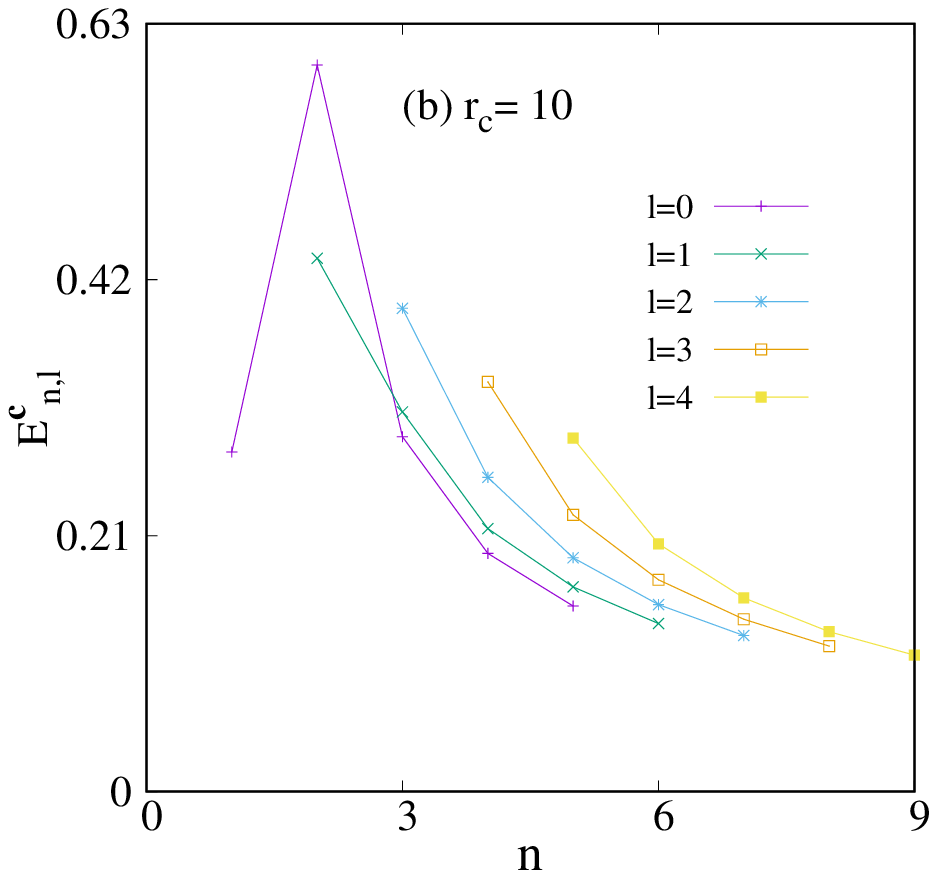}
\end{minipage}%
\caption{$E^{c}_{n,\ell}$ versus $n$ for $s,p,d,f,g$ states of CHA, at four $r_c$'s in panels 
(a)-(d). See text for details.}
\end{figure}
Next the focus is to understand the effect of state indices $n,\ell$ on $S^{c}$, which can help to interpret the influence of 
radial nodes on CPs. This can be discerned from Fig.~(3), where $S^c$ plots are displayed for $\ell=0-5$ states of CHA at $r_c$ 
values 0.1 (a), 10 (b), 20 (c), and $\infty$ (d), with $n$ up to 9. This first panel imprints that, at strong confinement regime 
($r_c=0.1$) $S^{c}_{n,\ell}$ grows with $n$, for a given $\ell$; however at a fixed $n$, it diminishes as $\ell$ goes up. That 
effectively implies, an increase in node in a system leads to growth of $S^{c}$ and consequently the broadening in CP.  Moreover, 
at fixed radial node condition, states with higher $n$ {\color{blue}possesses higher uncertainty in momentum.} For example, a 
2p state {\color{blue}imprints broader CP compared to a 1s state.} This leads to an important 
conclusion that the effect of confinement on a given state is completely governed by its principal quantum number and number of nodes.
In CHA, both $S_{c}$ as well as kinetic energy increase with $n$, leading to flattening of CP. Finally, the rate of release of Compton 
energy falls off as $n$ and nodes increase. In contrast, in panel (d), an exactly opposite trend of $S^{c}$ is witnessed, for the limiting 
case of FHA, where it reduces with increase in number of nodes. {\color{blue}In FHA due to the presence of accidental degeneracy (fixed $n$) the kinetic 
energy is independent of $\ell$. But its uncertainty decreases with rise in $\ell$. However, at a fixed $n$, $S^{c}$ increases with $\ell$.
It is a contradictory fact relative to CHA, where an enhancement in $S^{c}$ always stipulates the rise in uncertainty in both kinetic energy and 
momentum. Therefore, in FHA CP gets flattened with decrease in number of nodes. Similarly, at fixed $\ell$, $S^{c}$ abates with progress in $n$, explaining the 
increase in sharpness in CPs with rise in nodal structure. This, investigation clearly suggests that, in both FHA and CHA, CP can only be 
characterized and interpreted by invoking $S^{c}$. Therefore, this exploration establishes the role of $S^{c}$ as a descriptor in both CHA
and FHA. It is is a known fact that, while moving from cha TO FHA, behavorial pattern gets reversed. It can be explained by pointing out that, in 
FHA an increase in number of nodes shifts an orbital more away from the nucleus. However, in CHA rise in nodal structure squeezes the orbital more 
towards the centre.} The remaining two intermediate panels (b), (c) suggest the appearance of minima points in these plots, 
which, for a given state, moves towards right as $r_c$ enhances. Note that, in this figure, the minima are observable only for 
1s and 2p; for other states they become visible only at some sufficiently large $r_c$, which is not approached. This again 
concludes that, they act as bridge between CHA and FHA.

In order to revalidate the inferences drawn in Fig.~3, we have plotted $E^{c}_{n,\ell}$ of $s,p,d,f,g$ states (for $n$ up to 9) at 
four representative $r_{c}$'s namely $0.1, 10, 20, \infty$. Four panels (a)-(d) in Fig.~4 exhibit the same. As usual in all these 
four $r_{c}$, $E^{c}_{n,\ell}$ displays exact opposite behavior of $S^{c}_{n,\ell}$ of previous figure. As presumed, at $r_{c}=0.1$ 
in panel (a), $E^{c}_{n,\ell}$ decreases with rise in both $n$ and radial nodes respectively. On the contrary, panel (d) shows that, 
in case of FHA it increases for the same. This again reinforce the concept that (i) in CHA, pressure effect magnifies with growth
in $n$ and nodes (ii) rate of release in Compton energy reduces with $r_c$. In two middle panels (b) and (c), a maximum point is 
seen. Like $S^c_{n,\ell}$, this maximum point moves to right for greater $r_c$. It is also evident that, the qualitative 
behavior of $E^{c}_{n,\ell}$ and $\left\langle \frac{1}{2p} \right\rangle_{n,\ell}$ resemble each other.  

Finally, in panels (a)-(c) of Fig.~5, Shannon entropy density, ($-J(q)_{n,l} \ln J(q)_{n,l}$), has been plotted at three 
representative $r_{c}$ (0.1, 10, 20) respectively. Moving from left to right panels, the intensity of a given curve (for a 
fixed $n, \ell$ state) amplifies. Therefore, an increase in $r_c$ reduces the area under a given density curve. This was 
clearly manifested 
in first section of Table~II, where $S^{c}_{n,l}$ for a given state decayed with growth in $r_c$. At small $r_{c}$ region in panel 
(a), like CP, there appears several numbers of plateau in these curves. As usual, in a given state, number of such humps 
corresponds to the number of nodes. In moderate to large $r_{c}$ regions (b,c) the pattern of density curves changes. In 
left panel, for all six states, global maximum was located at $q=0$. But positions of such minima moves towards right direction 
in other two panels. Moreover, the number of humps in a given curve does not tally with number of nodes in that state. 

\begin{figure}                         
\begin{minipage}[c]{0.32\textwidth}\centering
\includegraphics[scale=0.55]{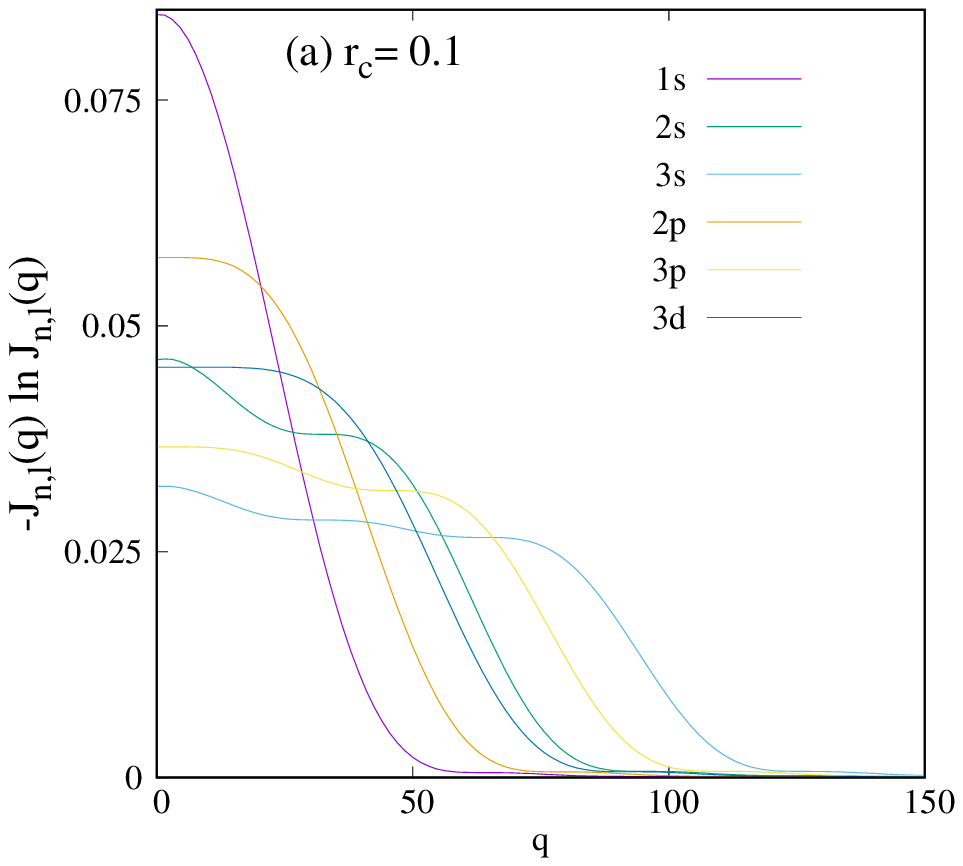}
\end{minipage}%
\vspace{1mm}
\begin{minipage}[c]{0.32\textwidth}\centering
\includegraphics[scale=0.55]{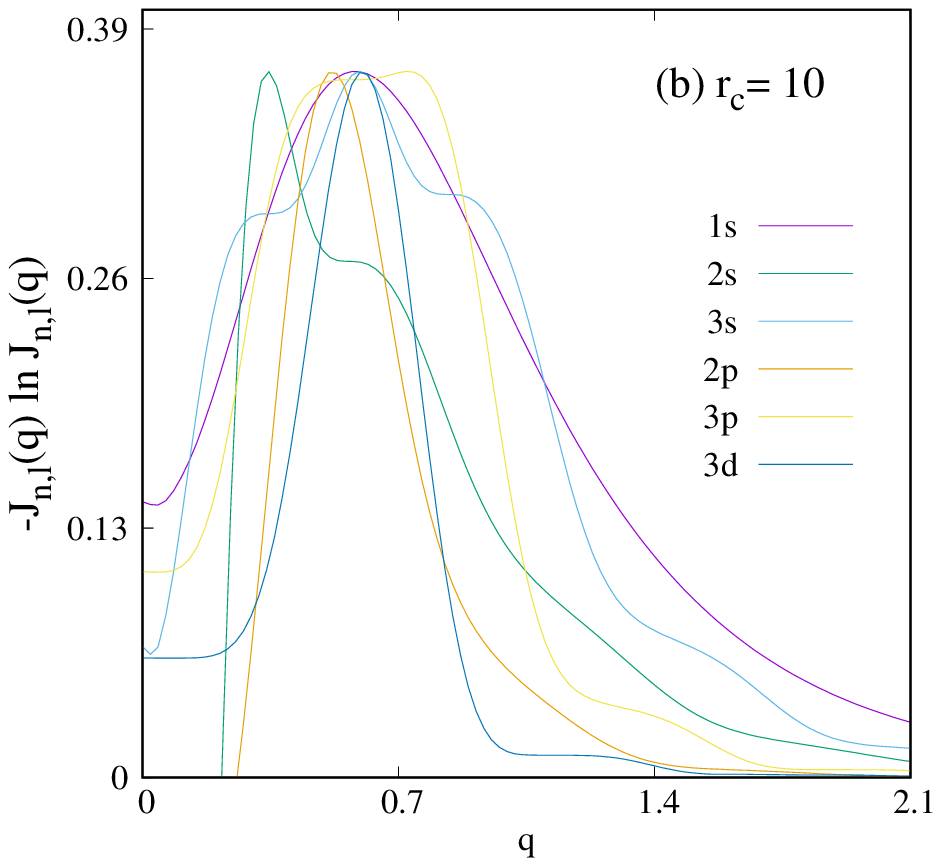}
\end{minipage}%
\vspace{1mm}
\begin{minipage}[c]{0.32\textwidth}\centering
\includegraphics[scale=0.55]{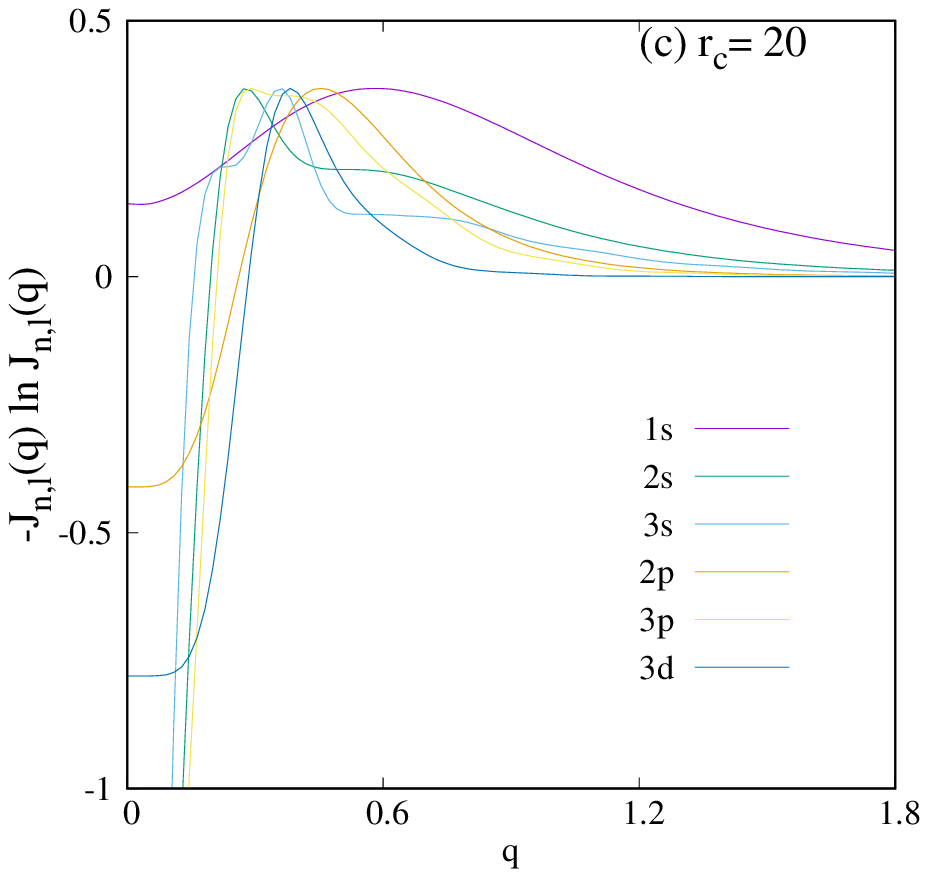}
\end{minipage}%
\caption{Shannon entropy density, ($-J_{n,\ell}(q) \ln J_{n,\ell}(q)$), for all $n \leq 3$ states in CHA at three $r_c=0.1, 10, 20,$ in 
panels (a)-(c) respectively. See text for details.}
\end{figure}

\subsection{Confined Hydrogen isoelectronic series}
The effect of $Z$ on confinement can easily be understood by analyzing Eqs.~(\ref{eq:7}), (\ref{eq:8}) and (\ref{eq:9}). Let us 
assume that, both $\hbar=1$ and $m=1$. Therefore, the moments becomes lead to, 
\begin{equation}
\left \langle \frac{1}{p} \right \rangle_{n,\ell} = \frac{1}{Z} \left \langle \frac{1}{p_{0}} \right \rangle_{n,\ell},  
\ \ \ \ \ \ \ \ \ \ \ \ \ \ 
\left \langle p^{m} \right \rangle_{n,\ell} = Z^{m} \left \langle p_{0}^{m} \right \rangle_{n,\ell}, \label{eq:10} 
\end{equation}
whereas the information entropies, in this case, turn out to be, 
\begin{equation}
S^{c}_{n,\ell}(1,Z,r_{c},q) = \frac{1}{2}\ln Z + S^{c}_{n,\ell}(1,1,Zr_{c},q_{0}), \ \ \ \ \ \ \ \ \  
E^{c}_{n,\ell}(1,Z,r_{c},q) = \frac{1}{Z}E^{c}_{n,\ell}(1,1,Zr_{c},q_{0})  \label{eq:11}.
\end{equation}
Equation~(\ref{eq:10}) suggests that, as one passes from $Z=2$ to 4, $\left \langle \frac{1}{p} \right \rangle_{n,\ell}$ lessens 
while $\left \langle p^{m} \right \rangle_{n,\ell}$ goes up. It is a known fact that, a rise in $\left \langle p^{m} \right 
\rangle_{n,\ell}$ is always accompanied by a drop in $\left \langle \frac{1}{p} \right \rangle_{n,\ell}$ and \emph{vice-versa}. 
Like other instances, here also $J_{n,\ell}(q=0)$ will provide a smooth monotonic decreasing curve, if plotted against 
$|\mathcal{E}|$ \cite{gadre79}. Moreover, it is seen from Eq.~(\ref{eq:11}) that, $S^{c}_{n,\ell}$ progresses and $E^{c}_{n,\ell}$ regresses 
with growth in $Z$. That means, the intensity of CP abates with rise in $Z$, which enhances the boundedness of electron. The 
ground-state $S^{c}, E^{c}$ of He$^{+}$, Li$^{2+}$, Be$^{3+}$, B$^{4+}$ are reported in Table~III at six distinct $r_{c}$ values, 
i.e., 0.1, 0.5, 1, 2, 10 and $\infty$. These results demonstrate the validity of expressions given in Eqs.~(\ref{eq:11}); in all 
four occasions, $S^{c}$ advances and $E^{c}$ declines with lengthening of box radius. Therefore, with relaxation in confinement, 
the rate of dissipation in kinetic energy magnifies. However, an enhancement in $Z$ value, inhibits the process.      

\begingroup           
\squeezetable
\begin{table}
\caption{$S^{c}$ and $E^{c}$ in the ground state of He$^{+}$, Li$^{2+}$, Be$^{2+}$, B$^{3+}$ ions at six $r_{c}$. 
See text for details. }
\centering
\begin{ruledtabular}
\begin{tabular}{l|llllllll}
 $r_{c}$ & \multicolumn{4}{c}{$S^{c}$} & \multicolumn{4}{c}{$E^{c}$} \\
\cline{2-5} \cline{6-9}
  &  $Z=2$ &  $Z=3$  & $Z=4$ & $Z=5$ & $Z=2$ &  $Z=3$  & $Z=4$ & $Z=5$ \\
\hline
0.1  & 1.69065 & 1.89338 & 2.03723 & 2.14880  & 0.007945  & 0.00793 & 0.00791  & 0.00789  \\ 
0.5  & 0.94337 & 1.14610 & 1.28994 & 1.40151  & 0.03881   & 0.03799 & 0.03694  & 0.03564    \\   
1    & 0.77344 & 0.97617 & 1.12001 & 1.23158  & 0.073875  & 0.06822 & 0.06093  & 0.05302    \\
2    & 0.75799 & 0.96082 & 1.10499 & 1.21666  & 0.12186   & 0.09142 & 0.06957  & 0.05572    \\
10   & 0.75796 & 0.96070 & 1.10458 & 1.21641  & 0.139345  & 0.09290 & 0.06967  & 0.05574    \\
$\infty$ & 0.757965 & 0.960698 & 1.104539 & 1.2161108 &  0.13926057 & 0.09284038 & 0.69630287 & 0.05570423 \\
\end{tabular}
\end{ruledtabular}
\end{table}
\endgroup

\section{Future and outlook}
Within the impulse approximation, CPs for confined H-like atoms, has been presented, for the first time.
The accuracy and reliability has been verified by calculating several momentum moments. Besides, $S^{c}$ 
and $E^{c}$ were also invoked in a novel way to analyze the CPs--their correctness suggests the impulse 
approximation to hold good in confined conditions. Such calculations are also performed in the respective unconfined 
systems. To the best of our knowledge, this is the first undertaking of such calculations. Moreover, they are 
found to play the role of good \emph{descriptors}, as they offer a proper interpretation about the 
boundedness of electron, and influence of radial node on a given state. As an offshoot, several interesting
analytical relations involving $S_{c}, E_{c}, \left \langle \frac{1}{p} \right \rangle, \langle p^{m} \rangle$, 
with $Z$, have been derived. It is observed that, the effect of $Z$ on CP's remains similar in both confined and 
free conditions. 

Experimental verification of these results is highly desirable. This will open up a new dimension in the high-pressure 
physics and chemistry. It is well known that, confined systems are generally, not exactly solvable;
and moreover, accurate calculation of energy and density may be quite demanding and challenging for conventional 
theoretical approximations. However, once CP is available experimentally, one can follow the usual route used in 
free condition to construct EMD, and subsequently wave function and energy. This may be an interesting recipe
to explain the bonding pattern, coordination number and reactivity in such stressed systems. Also, a theoretical 
exploration in many-electron systems would be quite helpful. Similar study for confined molecular systems may
provide vital insight about the effect of confinement in chemical bonding.       
\section{Acknowledgement}
Financial support from BRNS, India (sanction order: 58/14/03/2019-BRNS/10255) is gratefully acknowledged. NM thanks CSIR, 
New Delhi, India, for a Senior Research Associateship (Pool No. 9033A). {\color{blue} Critical constructive comments from 
an anonymous referee is greatly appreciated.} 

\section{Appendix~I: Onicescu energy calculation}
{\color{blue} At first let us work out the $\alpha$-order entropic moment, $\omega^{\alpha}$. It will provide an opportunity to calculate an 
arbitrary-order entropic moments. It can be derived as follows, 
\begin{equation}\label{eq:26}
\omega^{\alpha}= \int_{0}^{\infty} \left(\frac{8}{3\pi Z}\right)^{\alpha} \frac{1}
{\left[\left(\frac{q}{Z}\right)^{2}+1\right]^{3\alpha}}  \ \mathrm{d}q.
\end{equation} 
Assuming $\frac{q}{Z}=u$, this equation can be transformed in to, 
\begin{equation} \label{eq:27}
\omega^{\alpha}= \int_{0}^{\infty} \left(\frac{8}{3\pi}\right)^{\alpha} Z^{1-\alpha} \frac{1}{\left[u^{2}+1\right]^{3\alpha}} 
 \ \mathrm{d}u.
\end{equation} 
Further substitution of $u=\tan \theta$ in Eq.~(\ref{eq:27}) produces,
\begin{eqnarray} 
\omega^{\alpha} & = & \int_{0}^{\frac{\pi}{2}} \left(\frac{8}{3\pi}\right)^{\alpha} Z^{1-\alpha} \frac{\sec^{2}\theta}
{\left[\tan^{2}\theta+1\right]^{3\alpha}} \  \mathrm{d} \theta \\
& = & \left(\frac{8}{3\pi}\right)^{\alpha} Z^{1-\alpha} \int_{0}^{\frac{\pi}{2}}\cos^{6\alpha-2}\theta \ \mathrm{d}\theta.\label{eq:28}
\end{eqnarray}  
Now, making use of the standard integral, $\int_{0}^{\frac{\pi}{2}}\cos^{(\mu-1)} \theta \ \mathrm{d} \theta=2^{(\mu-2)}
B\left(\frac{\mu}{2},\frac{\mu}{2}\right)$ ($\mu=6\alpha-1$), in Eq.~(\ref{eq:27}), gives us the final result as, 
\begin{equation} \label{eq:29}
\omega^{\alpha}= \left(\frac{8}{3\pi}\right)^{\alpha} Z^{1-\alpha} \  2^{(6\alpha-3)}B\left(\frac{(6\alpha-1)}{2},\frac{(6\alpha-1)}
{2}\right). 
\end{equation} 
It is needless to mention that, at $\alpha=2$, this entropic moment in Eq.~(\ref{eq:29}) reduces to $E^{c}_{n,\ell}$. Therefore, 
\begin{equation} \label{eq:30}
E^{c}_{1,0} = \omega^{2}  = \left(\frac{8}{3\pi}\right)^{2} \frac{1}{Z} B\left(\frac{11}{2},\frac{11}{2}\right) 
= \left(\frac{8}{3\pi}\right)^{2} \frac{1}{Z} \ \frac{\Gamma (\frac{11}{2}) \Gamma (\frac{11}{2})}{\Gamma (11)} = 
\left[\frac{7}{8\pi Z}\right]
\end{equation}}

\section{Appendix~II: Shannon entropy calculation}
{\color{blue}In a similar fashion, $S^{c}_{n,l}$ for $1s$ state will take the form, 
\begin{equation}\label{eq:31}
S^{c}_{1,0} = - \int_{0}^{\infty} \left(\frac{8}{3\pi Z}\right) \frac{1}{\left[\left(\frac{q}{Z}\right)^{2}+1\right]^{3}} \ 
\ln \left[\left(\frac{8}{3\pi Z}\right) \frac{1}{\left[\left(\frac{q}{Z}\right)^{2}+1\right]^{3}}\right] \mathrm{d}q. 
\end{equation} 
Once again putting $\frac{q}{Z}=u$ and $u=\tan \theta$, Eq.~(\ref{eq:31}) can be recast in to,
\begin{equation}
\begin{aligned}
S^{c}_{1,0} & =  \left[\ln \left(\frac{3\pi}{8}\right)+ \ln Z\right]\left(\frac{8}{3\pi}\right) \int_{0}^{\frac{\pi}{2}} \cos^{4} 
 \theta \ \mathrm{d} \theta
-\frac{16}{\pi} \int_{0}^{\frac{\pi}{2}} \cos^{4} \theta \ \ln \ \cos \theta \ \mathrm{d} \theta, \\ 
S^{c}_{1,0} & =  \left[\ln \left(\frac{3\pi}{8}\right)+ \ln Z\right]\left(\frac{8}{3\pi}\right) A_{1}-\frac{16}{\pi} A_{2}. 
 \label{eq:32}
\end{aligned}
\end{equation}
Now invoking the following standard integrals \cite{gradshteyn15}, 
\begin{equation}
\begin{aligned}
\int_{0}^{\frac{\pi}{2}}\cos^{2m} \theta \mathrm{d} \theta & =  \frac{(2m-1)!!}{(2m)!!}\frac{\pi}{2},   \\ 
\int_{0}^{\frac{\pi}{2}} \cos^{2k} \theta \ \ln \cos \theta \mathrm{d} \theta & =  -\frac{(2k-1)!!}{2^{k}k!} \ \frac{\pi}{2}  
\left[ \ln 2 + \sum_{m_{1}=1}^{k} \frac{(-1)^{m_{1}}}{m_{1}}\right],
\end{aligned}
\end{equation}
we obtain $A_1, A_2$ as, 
\begin{equation}
\begin{aligned}
A_{1} & =  \int_{0}^{\frac{\pi}{2}} \cos^{4} \theta \mathrm{d} \theta   =   \frac{3\pi}{16},  \\
A_{2} & =  \int_{0}^{\frac{\pi}{2}} \cos^{4} \theta \ \ln \cos \theta \mathrm{d} \theta  =  -\frac{3\pi}{16} \left[\ln 2 -
\frac{7}{12}\right]. 
\end{aligned}
\end{equation}
Finally inserting the values of $A_{1}$ and $A_{2}$ in Eq.~(\ref{eq:32}) one gets,
\begin{equation}\label{eq:33}
S^{c}_{1,0}=\frac{1}{2} \ln 24 \pi + \frac{1}{2} \ln Z- \frac{7}{4}.
\end{equation}}

\end{document}